\documentclass[iop,natbib]{emulateapj}
\usepackage{amsmath,amssymb}
\usepackage{epstopdf}

\newcommand{\mytitle}{Non-relativistic perpendicular shocks modeling young supernova remnants: nonstationary dynamics and particle acceleration at forward and reverse shocks}

\newcommand{\rev}{\bf }

\usepackage{natbib}
\usepackage{comment}

\shorttitle{Non-relativistic perpendicular shocks}
\shortauthors{Wieland et al.}

\begin{document}

\title{\mytitle}

\author{Volkmar Wieland\altaffilmark{1}, Martin Pohl\altaffilmark{1,2,5}, Jacek Niemiec\altaffilmark{3}, Iman Rafighi\altaffilmark{1}, Ken-Ichi Nishikawa\altaffilmark{4}}
\email{marpohl@uni-potsdam.de}
\altaffiltext{1}{Institute of Physics and Astronomy, University of Potsdam, 14476 Potsdam, Germany}
\altaffiltext{2}{DESY, 15738 Zeuthen, Germany}
\altaffiltext{3}{Instytut Fizyki J\c{a}drowej PAN, ul. Radzikowskiego 152, 31-342 Krak\'{o}w, Poland}
\altaffiltext{4}{Department of Physics, University of Alabama in Huntsville, Huntsville, AL 35899, USA}
\altaffiltext{5}{Corresponding author}

\begin{abstract}
For parameters that are applicable to the conditions at young supernova remnants, we present results of 2D3V particle-in-cell simulations of a non-relativistic plasma shock with a large-scale perpendicular magnetic field {inclined at $45^{\rm o}$ angle to the simulation plane to approximate 3D physics}. We developed {an improved} clean setup that uses the collision of two plasma slabs with different density and velocity, leading to the development of two distinctive shocks and a contact discontinuity. 
The shock formation is mediated  by Weibel-type filamentation instabilities that generate magnetic turbulence. 
Cyclic reformation is observed in both shocks with similar period, for which we note global variations on account of shock rippling and local variations arising from turbulent current filaments. 
{The shock rippling occurs on spatial and temporal scales given by gyro-motions of shock-reflected ions.}
The drift motion of electrons and ions is not a gradient drift, but commensurates with $\mathbf{E}\times\mathbf{B}$ drift. We observe a stable suprathermal tail in the ion spectra, but no electron acceleration because the amplitude of Buneman modes in the shock foot is insufficient for trapping relativistic electrons. We see no evidence of turbulent reconnection. A comparison with other {2D} simulation results suggests that the plasma beta {and the ion-to-electron mass ratio are} not decisive for efficient electron acceleration, {but pre-acceleration efficacy might be reduced with respect to the 2D results once three-dimensional effects are fully accounted for}. Other {microphysical} factors 
{may also} be at play to limit the amplitude of Buneman waves or prevent return of electrons to the foot region.
\end{abstract}


\section{Introduction}\label{introduction}
Collisionless shocks in space are sites of efficient particle acceleration. While direct in-situ studies of these shocks are possible only in interplanetary space, understanding the properties of high Mach-number shocks in
supernova remnants (SNRs) is highly desirable, because SNR are suspected to supply a significant fraction of cosmic rays, and emission of freshly accelerated particles has been observed for many years \citep{reynolds}. The likely dominant acceleration process is
diffusive shock acceleration (DSA) or first-order Fermi acceleration \citep{fermi}. Charged particles scatter off magnetic inhomogeneities, e.g., in the form of magnetohydrodynamic (MHD) waves, in the upstream and downstream regions of the shock, which isotropizes their distribution function in each region. If particles have enough energy to cross the shock front, i.e., their mean free path is large enough to ``see'' the shock as a sharp discontinuity, they systematically gain energy with each cycle of shock crossing, and the relation between the probabilities of escape and of return to the shock determines what spectrum the particles assume \citep{blandford_1987}.

Relevant for the process are the structure of the shocks, the electromagnetic field amplitudes at them, and the local pre-acceleration processes that separate particles from the quasithermal bulk. The shock structure is typically driven by ions and, therefore, has ion length scales. Electrons have small plasma and gyration scales compared with ions, and so electron pre-acceleration is a particularly interesting problem.

Here we study perpendicular shocks whose structure is driven by ion reflection, leading to a steep density ramp and various instabilities operating in the foot region ahead of the ramp \citep{2005SSRv..118..161B,treu_jaro3,treumann2009}. Depending on the Alfv\'enic Mach number, $M_\mathrm{A}$, quasi-standing whistler waves may be found in the foot region \citep{2007GeoRL..3414109H}, 
and accelerate electrons \citep{riqspit_2011}. The large-scale motion of ions can lead to shock surfing acceleration (SSA) \citep{ssa_1973} or shock drift acceleration (SDA) \citep{sda_1989}. In their original design these two process would not significantly affect electrons \citep{treu_jaro5}, but see \citet{guo_sironi_2014}. Instead, appropriate instabilities, solitary structures or similar are required, an example of which is the Buneman instability between reflected ions and incoming electrons. Strong electron heating {and accelaration} would result in the foot region \citep{2000ApJ...543L..67S}, possibly followed by secondary {accelerations} through adiabatic {processes} or the ion-acoustic instability \citep{kato_2010,amahosh_2012,2013PhRvL.111u5003M}.

Studying electron acceleration requires that electron scales be resolved, and so we conduct particle-in-cell (PIC) simulations in 2D3V configuration, i.e. allowing gradients in 2 dimensions but following all 3 vector components. Hybrid simulations \citep[e.g.,][]{capspit_2014a} permit studying the behavior on longer time scales and large spatial scales, but provide no information on electron dynamics. Having investigated unmagnetized shocks and weakly magnetized strictly parallel shocks before \citep[hereafter N2012]{niemiec_2012},  
we here report on simulations of strictly perpendicular ($\theta_{Bn}=90^{\circ}$) shocks, leaving results for oblique shocks to a future publication.

We concentrate on the parameter regime typical of young SNRs, as opposed to, e.g., heliospheric conditions and {low Mach number nonrelativistic shocks \citep[e.g.,][]{umeda2008,riqspit_2011,guo_sironi_2014}.}
{We consider large sonic (cold plasmas with $\beta_e \ll 1$)} and Alfv\'{e}nic Mach numbers, $M_\mathrm{s}$ and $M_\mathrm{A}$, and sub-relativistic plasma collision velocities allowing for Weibel-type (filamentation) instabilities to occur \citep{kato2008,kato_2010,niemiec_2012}.
Our simulations thus complement the 2-D PIC simulations of \citet{2013PhRvL.111u5003M} and \citet{2015Sci...347..974M} who have covered the parameter range $\beta_e\simeq 0.5$ and both sonic and Alfv\'enic Mach number around 40. 
In contrast to all earlier studies of high Mach number perpendicular shocks that use either in-plane or out-of-plane configurations of the homogeneous magnetic field {\rev and often find a different efficiency of, e.g., electron heating by electrostatic modes \citep{amahosh_2009}, we set our regular magnetic field component at an angle of $45^{\rm{o}}$ to the simulation plane. 
We expect that such setup will better approximate the physics of the fully three-dimensional systems.} 
We follow the shock evolution for 20 ion cyclotron times, $t\,\Omega_{i} =20$, considerably longer than other published PIC studies. We have also introduced a setup that minimizes artificial electromagnetic transients during shock launching, that may influence the upstream medium.

{In numerical simulations, collisionless shocks can be initiated in a number of ways. The most widely used are the injection method \citep{burgess}, the flow-flow method \citep{1992JGR....9714801O}, the relaxation method \citep{leroy1981,leroy1982}, and the magnetic-piston method \citep{1992PhFlB...4.3533L}. The injection method uses a plasma beam that is reflected off a conducting wall. The reflected particle beam interacts with the incoming plasma, and a shock is created. Implicitely this method assumes an infinitely sharp contact discontinuity (CD), whereas in a collisionless plasma the CD has a finite width and internal structure. 
In the flow-flow method two counterstreaming plasma beams are continuously injected at sides of the computational box and couple to form in time a system of two shocks separated by a CD. 
This setup offers more freedom in the choice of the physical parameters for the colliding plasmas. It is also physically more accurate than the injection method
as it avoids the assumption of an infinitely sharp CD. The relaxation method uses a simulation box filled with plasma that is separated by a discontinuity into two uniform plasma slabs that are supposed to initially satisfy the shock jump conditions \citep[see also][]{2006EP&S...58E..41U}. The magnetic piston method, at last, applies an external current pulse that induces an electromagnetic field transient which propagates in the plasma to develop into a shock \citep[for a more detailed account of the shock excitation methods see, e.g.,][]{2003LNP...615...54L}.}


{In our simulations, we use a flow-flow method of shock excitation that allows us to investigate the dynamics of both forward and reverse shock at the same time. 
The asymmetric slab-collision setup used in Niemiec et al. (2012) is further developed in the present study} to include a new setup for perpendicular shocks that avoids having sharp gradients in the motional electric fields with opposite sign at the CD. Such gradients work as an artificial dipole antenna and thereby emit a strong electromagnetic pulse whose presence in the system may limit the veracity of the simulation in the initial, as well as non-linear, stage of the system evolution.

We describe 
{updates to our} simulation model and the setup in Section \ref{setup}. The results of the simulation are presented in Section \ref{results}. A summary and discussion conclude the paper in Section \ref{summary}.

\section{Simulation Setup}\label{setup}
\subsection{New Setup for Perpendicular Shocks}\label{new_setup}
The numerical grid is initially filled with two plasma slabs of uniform density that are separated by a void and move towards each other. Once they collide, a system of two shocks and the CD is formed. Note, that in contrast to the injection and the relaxation methods, the CD is self-consistently developed and not initially assumed.

\begin{figure}[htb]
\includegraphics[width=\linewidth]{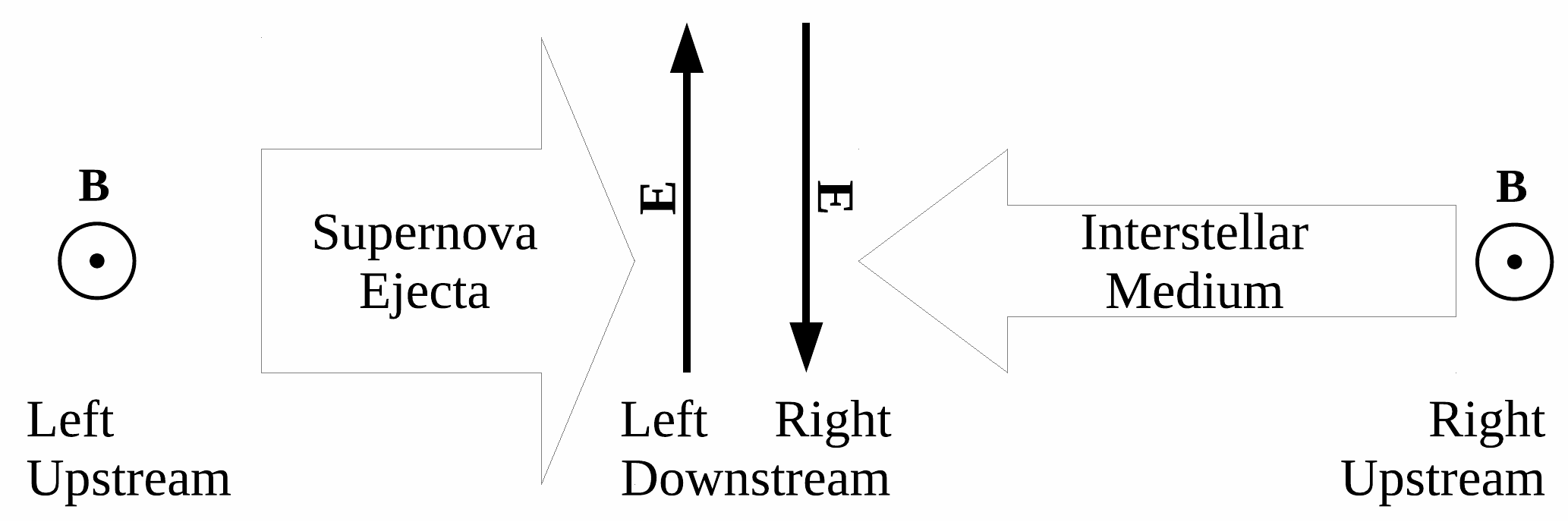}
\caption{Illustration of the setup for a perpendicular shock showing the directions of motion of the two plasma slabs and the motional electric fields for a magnetic field oriented out of the plane of the figure.}\label{thatmpi_setup}
\end{figure}

To set up a magnetized plasma system one may establish in the entire simulation box a homogeneous magnetic field $\mathbf{B}_0$ perpendicular to the streaming direction of the plasma beams. The magnetic field is meant to be frozen into the moving plasma, i.e. in the rest frame of the plasma $\mathbf{E}'=0$ and $\mathbf{B}'$ are the electric and magnetic field. In the simulation frame, nonrelativistic Lorentz transformations yield $\mathbf{E}=-\mathbf{v}\times\mathbf{B}'$ and $\mathbf{B}=\mathbf{B}'$, where $\mathbf{v}$ is the streaming velocity of the plasma slabs. In our simulation the two plasmas have streaming velocities $\mathbf{v}_\mathrm{L}=v_{\mathrm{L},x}\,\mathbf{\hat{x}}$ and $\mathbf{v}_\mathrm{R}=v_{\mathrm{R},x}\,\mathbf{\hat{x}}$, the indices L and R referring to the initial position of the plasmas on the \textit{left} and \textit{right} side of the simulation box, respectively, and the magnetic field is aligned in $y$-$z$-direction, which leads to
\begingroup
\renewcommand*{\arraystretch}{1.2}
\begin{equation}
\left(\begin{array}{c}
E_{\mathrm{L},x} \\ E_{\mathrm{L},y} \\ E_{\mathrm{L},z}
\end{array}\right)
=
\left(\begin{array}{c}
0 \\ v_{\mathrm{L},x}\,B_{\mathrm{L},z} \\ -v_{\mathrm{L},x}\,B_{\mathrm{L},y}
\end{array}\right)
\end{equation}
and
\begin{equation}
\left(\begin{array}{c}
E_{\mathrm{R},x} \\ E_{\mathrm{R},y} \\ E_{\mathrm{R},z}
\end{array}\right)
=
\left(\begin{array}{c}
0 \\ v_{\mathrm{R},x}\,B_{\mathrm{R},z} \\ -v_{\mathrm{R},x}\,B_{\mathrm{R},y}
\end{array}\right).
\end{equation}
\endgroup
In the simulations described here, $\mathbf{B}_\mathrm{L}=\mathbf{B}_\mathrm{R}$.
Since $v_{\mathrm{L},x}$ and $v_{\mathrm{R},x}$ have opposing signs, the motional electric field has opposing signs in the two plasmas, which is illustrated in Figure~\ref{thatmpi_setup}. Without further modification, this setup would lead to a large value of $\nabla\times\mathbf{E}$ {at the edges of the plasma slabs} in the middle of the simulation box, that through the corresponding $\partial\mathbf{B}/\partial t$ would induce an electromagnetic transient that may limit the veracity of the simulation.
 
{We developed a setup that avoids this artificial antenna effect by implementing a transition zone 
between the two slabs. A spatial gradient is imposed in the perpendicular magnetic-field components $B_y$ and $B_z$ 
in front layers of the colliding plasma slabs that tapers the field off until it vanishes at the interface to the plasma-free area, which can initially separate the slabs.
A corresponding tapering of the motional electric field results naturally. The nonzero $\nabla\times\mathbf{B}$ is compensated by a current sheet in which ions drift relative to the electrons.} A detailed mathematical description and illustration of this new setup can be found in Appendix~\ref{setup_details}.

Figure \ref{compare} in Appendix \ref{setup_details} compares the stability in particle and field density of the new setup with the conventional scenario of a jumping motional electric field, $E_y$, on account of a constant perpendicular magnetic field, $B_z$. The new setup is very stable over many time steps, whereas for the standard setup with constant magnetic field one can clearly see a transient in the electric field, which is emitted in the middle of the simulation box and eventually perturbs the magnetic field.

\subsection{Simulation Parameters}\label{parameters}
For ease of comparison, we basically used the same numerical parameters as for simulation run M1 in N2012. The parameters are chosen such that our simulation results may be applied to plasma shocks formed at young supernova remnants.
The two plasma slabs are composed of equal numbers of electrons and ions, which are initialized at the same location in order to make the system initially neutral. The two slabs have different densities with a density ratio 10, and hence their plasma frequencies differ by a factor $\sqrt{10}\simeq 3.1$. We use ten particles per cell per particle species for both plasma slabs and assign statistical weights to the tenuous-plasma particles to establish the intended density ratio. In order to resolve the characteristic length scales of both electrons and ions in our simulations, we choose a reduced {ion-to-electron} mass ratio of $m_i/m_e=50$.

\begin{table}[htb]
\caption{Basic parameters of the double-shock simulation and derived shock properties.}
\begin{center}
\begin{tabular}{lrr}\hline
 & Left& Right\\
 & Dense plasma& Dilute plasma \\
 & Reverse shock& Forward shock\\
\hline
Skin length & $\lambda_{se,\mathrm L}$& $\lambda_{se,\mathrm R}$\\
 & 7.9\,$\Delta$ & 25\,$\Delta$ \\
 \hline
Thermal speed & & \\
$v_{e,\mathrm{th}}$ & 0.002\,c & 0.002\,c \\
\hline
Streaming speed & $v_{\mathrm{L},x}$&$v_{\mathrm{R},x}$ \\
 & 0.0354\,c &\hphantom{bla} -0.354\,c \\
 \hline
Alfv\'en speed & $v_\mathrm{A, L}$ & $v_\mathrm{A, L}$ \\
 &\hphantom{bla} 0.00447\,c & 0.0142\,c \\
\hline
Shock speed & $v_\mathrm{sh,L}$ & $v_\mathrm{sh,R}$ \\
in upstream frame \hphantom{bla}& -0.127\,c & 0.39\,c \\
\hline
Mach numbers & & \\
$M_\mathrm{A}$ & 28.5 & 27.6 \\  
$M_\mathrm{s}$ & 252 & 755 \\ \hline
\end{tabular}
\end{center}
\label{params}
\end{table}

The main simulation parameters are summarized in Table~\ref{params}.
In our simulation frame, which is the center-of-momentum frame, the \textit{dense} plasma moves to the right, while the \textit{tenuous} plasma moves to the left. The two plasmas collide at a relative speed of $v_{\mathrm{rel}}=0.38\,c$ and {in time} form a double-shock structure with a contact discontinuity (CD) separating the downstream regions of the two shocks. 
On account of the similarity with dense ejecta moving into dilute ambient medium in a SNR, we designate the right shock in the low-density plasma as the forward shock and the left shock as the reverse shock. In time, the CD starts moving in the simulation frame with $v_{\mathrm{CD}}=-0.06\,c$ on account of the system being in momentum balance but not in ram-pressure balance.

Our simulations are performed using a 2D3V model, i.e., we restrict all particles to two spatial dimensions while keeping all three components of their velocities. Since the large-scale magnetic field bends the particle trajectories out of the simulation plane, particles have three degrees of freedom, and the non-relativistic adiabatic index $\Gamma=5/3$. 
The electrons and ions of both plasmas are initially in thermal equilibrium and cold, thus permitting large sonic Mach numbers as expected for SNR. Initially homogeneous magnetic field is aligned perpendicular to the plasma flow 
{and lies in the $y-z$ plane making an angle $\phi=45^{\rm{o}}$ with the $y-$axis,}
i.e., $\mathbf{B}=B_0\,(0,1,1)/\sqrt{2}$. The strength of the magnetic field can also be expressed by the electron cyclotron frequency, $\Omega_{e}=eB_0/m_e$, whose ratio to the electron plasma frequency of the dense plasma, $\omega_{pe,L}=\sqrt{n_{e,L}e^2/\epsilon_0m_e}$, is $\Omega_{e}/\omega_{pe,L}=0.032$. Here $n_{e,L}$ is the electron density of the dense plasma, {$e$ is the electric charge,} and $\epsilon_0$ the permittivity of vacuum. 

{The sonic and Alfv\'enic Mach numbers given in Table~\ref{params} are calculated as the ratios of the shock velocities in the upstream reference frames to the sound and Alfv\'en speeds in these frames.}
The sonic Mach numbers are typical for SNRs expanding into a medium of temperature around a few thousand Kelvin. 
{The Alfv\'enic Mach numbers of our two shocks are considerably lower than minimum Mach numbers of $M_A \approx 180$ estimated for the real mass ratio $m_i=1836\,m_e$ for shocks in young SNRs to efficiently pre-accelerate electrons \citep[see Eq. 8 in][]{amahosh_2012}. However, for the reduced mass ratio employed in our simulations we can sample the same physics already at M$_A\geq 16$. The Alfv\'enic Mach numbers in our simulations can thus well represent the conditions at young SNR shocks 
expanding into a weakly magnetized medium.}

The spatial dimensions in our simulation and all figures are given in terms of the electron skin length of the dense plasma $\lambda_{se}\equiv\lambda_{se,L}=c/\omega_{pe,L}=7.9\,\Delta$, where $\Delta$ is the size of the grid cells. The temporal dimensions are given in terms of the inverse of the upstream ion Larmor frequency $\Omega_{i}^{-1}=1582.3\,\omega_{pe,L}^{-1}$. The simulation time is $T=20\,\Omega_{i}^{-1}=31645.5\,\omega_{pe,L}^{-1}=4472.3\,\omega_{pi,L}^{-1}=1000\,\Omega_{e}^{-1}$, which is longer by a factor 4 and 2 than the simulations described in \citet{kato_2010} and \citet{guo_sironi_2014}, respectively. 

The transverse size of the simulation box is $L_y=324.1\,\lambda_{se}=42.9\,\lambda_{si,L}$ with periodic boundary conditions. 
{Radiative boundary conditions are imposed in $x$-direction.}
We use a so-called moving injector {method, in which} particles are continuously injected in a small injection layer at the outer edges of the two plasma slabs. 
The injection layer retreats from the collision zone as far as necessary to {keep all streaming or reflected} particles and fields within the collision region, 
but it stays as close as possible such that newly injected particles do not have to travel a long distance without any interaction. The simulation box is thus permitted to grow in $x$-direction, {reaching} the final size of $L_x=9493.7\,\lambda_{se}$ {at the end of the simulation.} 

The simulation was performed with a modified version of the TRISTAN code \citep{tristan}, which was adapted to work in 2D3V and parallelized using MPI \citep{niemiec_2008}. Other modifications include a fourth-order finite-difference time-domain (FDTD) field-pusher with a weak Friedman filter \citep{greenwood,friedman} that efficiently filters numerical Cerenkov radiation, and the pusher proposed in \citet{vay} that achieves better energy conservation and less numerical self-heating.

\section{Simulation Results}\label{results}
\subsection{Early-stage Evolution}\label{early_stage}
We remind the reader that the right shock in the dilute plasma is also referred to as the forward shock and the left shock in the dense medium is the reverse shock. Also recall that as length scale we use the electron skin length, $\lambda_\mathrm{se}$, of the far-upstream dense plasma in the left part of the simulation box. Downstream of the reverse shock the true skin length is only $\lambda_\mathrm{se}/2$, and less than that at the density overshoot. In the tenuous plasma on the right the true skin length is always $3.1$ times that in the dense plasma on account of factor-$10$ density ratio. 

\begin{figure}[htb!]
\centering
\includegraphics[width=\linewidth]{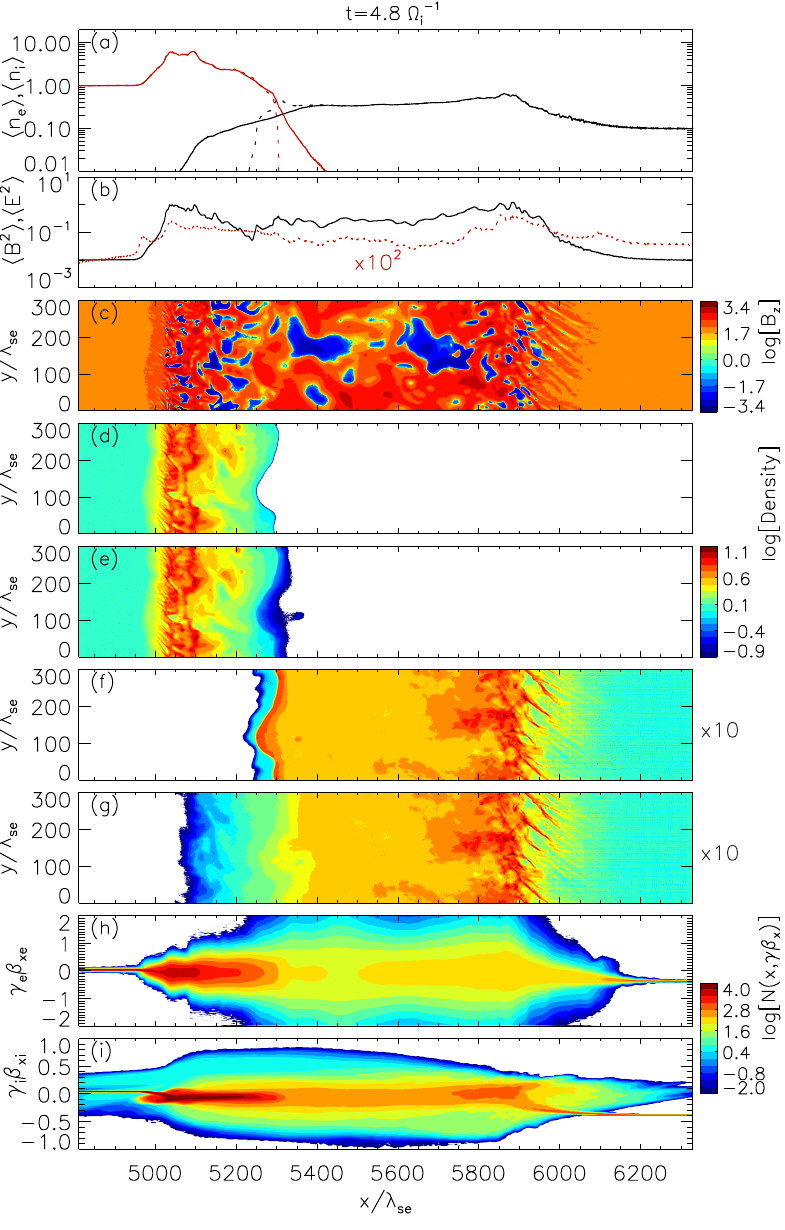}
\caption{Structure of the plasma-collision region at time $t=4.8\,\Omega_{i}^{-1}$. Displayed are the profiles of (a) the average particle-number density normalized to the far-upstream density of the dense plasma (red lines: dense plasma, black lines: tenuous plasma; solid lines: ions, {dashed} lines: electrons), (b) profiles of the average magnetic (solid line) and electric (red dotted line, times factor 100) energy density in simulation units, (c) the amplitude of the magnetic field $B_z$ (in sign-preserving logarithmic scale as $\mathrm{sgn}(B_z)\,(2+\log\left[\max(10^{-2},|B_z|)\right])$), the density of dense-plasma electrons (d), dense-plasma ions (e), tenuous-plasma electrons (f), and tenuous-plasma ions (g), all normalized to their far-upstream values, and the longitudinal phase-space distribution of electrons (h) and ions (i).}
\label{plot_double_shock}
\end{figure}

In this section we present the structure of the two evolving shocks after $T\,\Omega_i =4.8$, which is comparable to the total time covered in the simulations of \citet{umeda2008}, \citet{kato_2010}, and \citet{amahosh_2012}. A global picture is offered in Figure~\ref{plot_double_shock}, where we show profiles of the transversely-averaged particle densities and field energy densities, $p_x$ vs $x$ phase-space information, and 2-D distributions of particle density and the magnetic-field component $B_z$.

\begin{figure}[htb!]
\centering
\includegraphics[width=\linewidth]{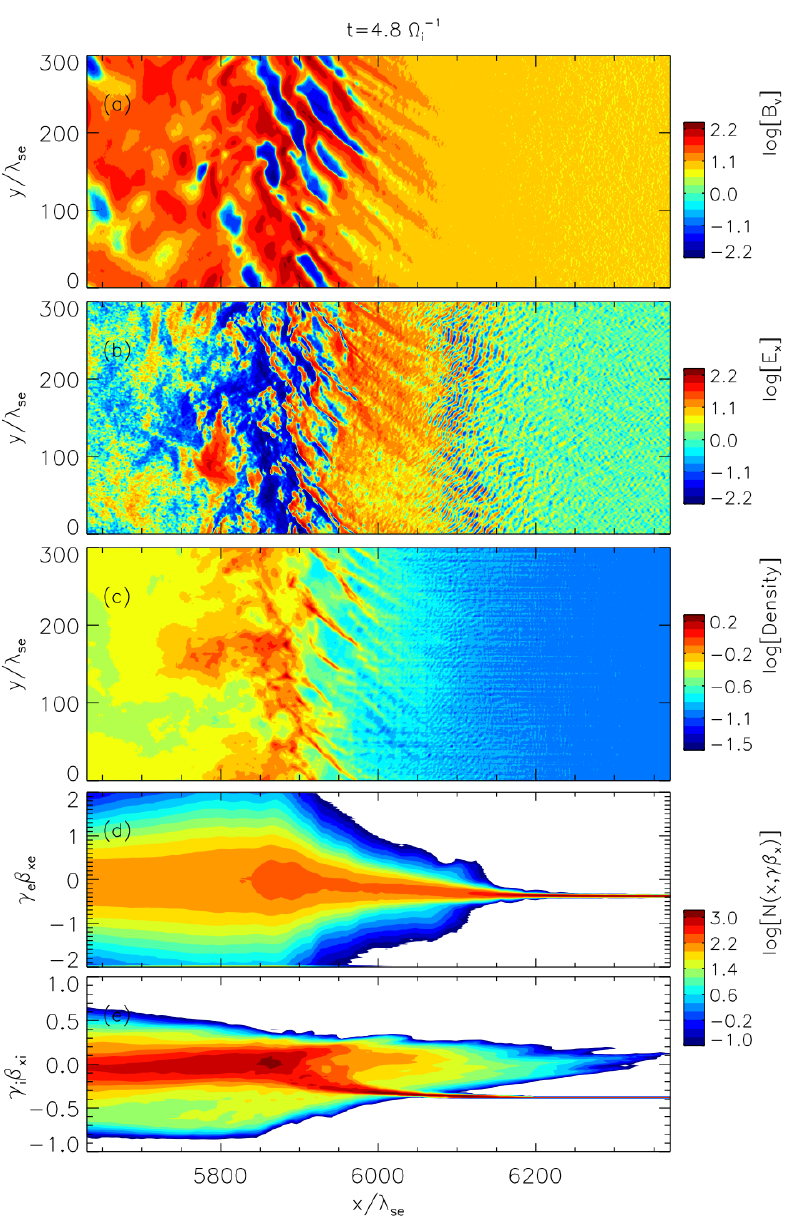}
\caption{Structure of the forward shock at time $t=4.8\,\Omega_{i}^{-1}$. Displayed are the distributions of the magnetic-field component $B_y$) (a), the electric-field component $E_x$ (b), the density of electrons (c), and the longitudinal phase-space distribution of electrons (d) and ions (e).}
\label{fw_shock}
\end{figure}

\begin{figure}[htb!]
\centering
\includegraphics[width=\linewidth]{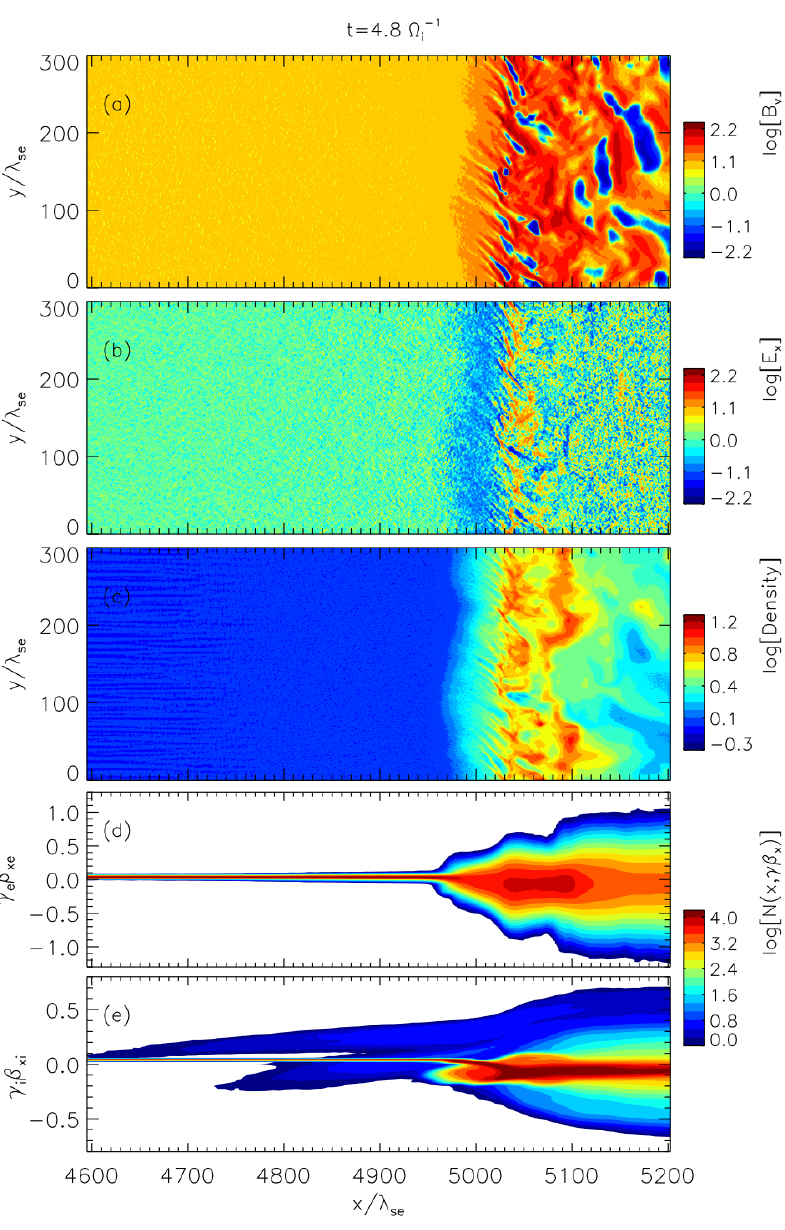}
\caption{Structure of the reverse shock at time $t=4.8\,\Omega_{i}^{-1}$. Displayed are the distributions of the magnetic-field component $B_y$ (a), the electric-field component $E_x$ (b), the density of electrons (c), and the longitudinal phase-space distribution of electrons (d) and ions (e).}
\label{rv_shock}
\end{figure}

The CD, now clearly visible at $x\approx 5300\,\lambda_\mathrm{se}$, moves with the predicted speed of $v_{\mathrm{CD},x}=-0.06\,c$. The forward shock is visible at about $x\approx 5900\,\lambda_\mathrm{se}$, and the reverse shock is found near $x\approx 5000\,\lambda_\mathrm{se}$.
It is obvious that the two shocks have not propagated very far from the CD, and so the system may not yet be in a statistical equilibrium. Nevertheless, the salient features of collisionless shocks are already visible. 
Noting that the 2-D distribution of ions very closely follows that of the electrons, we only show the electron density in the blow-up of the display for the forward shock (Figure~\ref{fw_shock}) and the reverse shock (Figure~\ref{rv_shock}).

To be noted from Figure~\ref{fw_shock} is the presence of shock-reflected ions which in a
supercritical perpendicular shock 
may lead to the growth of Buneman modes in the foot of the shock, which {can} subsequently heat {and accelerate} the incoming electrons. A Fourier analysis of $E_x$ indeed reveals Buneman waves at $k_\perp\,\lambda_\mathrm{se}\simeq 0$ and $k_\parallel\,\lambda_\mathrm{s,e}\simeq 0.8$. Figure~\ref{fw_shock} indicates that the drift velocity between reflected ions and incoming electrons near $x=6150\,\lambda_\mathrm{se}$ is $v_\mathrm{rel}\simeq 0.4\,c$. We expect the growth of Buneman modes at 
\begin{equation}
k_\parallel\simeq \frac{c}{v_\mathrm{rel}}\,\frac{1}{\lambda_\mathrm{se,local}}\simeq 
2.5\,\frac{1}{3.1\,\lambda_\mathrm{se}}\simeq 0.8\,\frac{1}{\lambda_\mathrm{se}}\ ,
\label{eq0}
\end{equation}
which is exactly what is observed. In Equation~\ref{eq0} we have used the local electron skinlength, $\lambda_\mathrm{se,local}=\lambda_{se,\mathrm R}$, which is $\sqrt{10}$ times that of the dense plasma, $\lambda_\mathrm{se}$.

\begin{figure}[htb!]
\centering
\includegraphics[width=\linewidth]{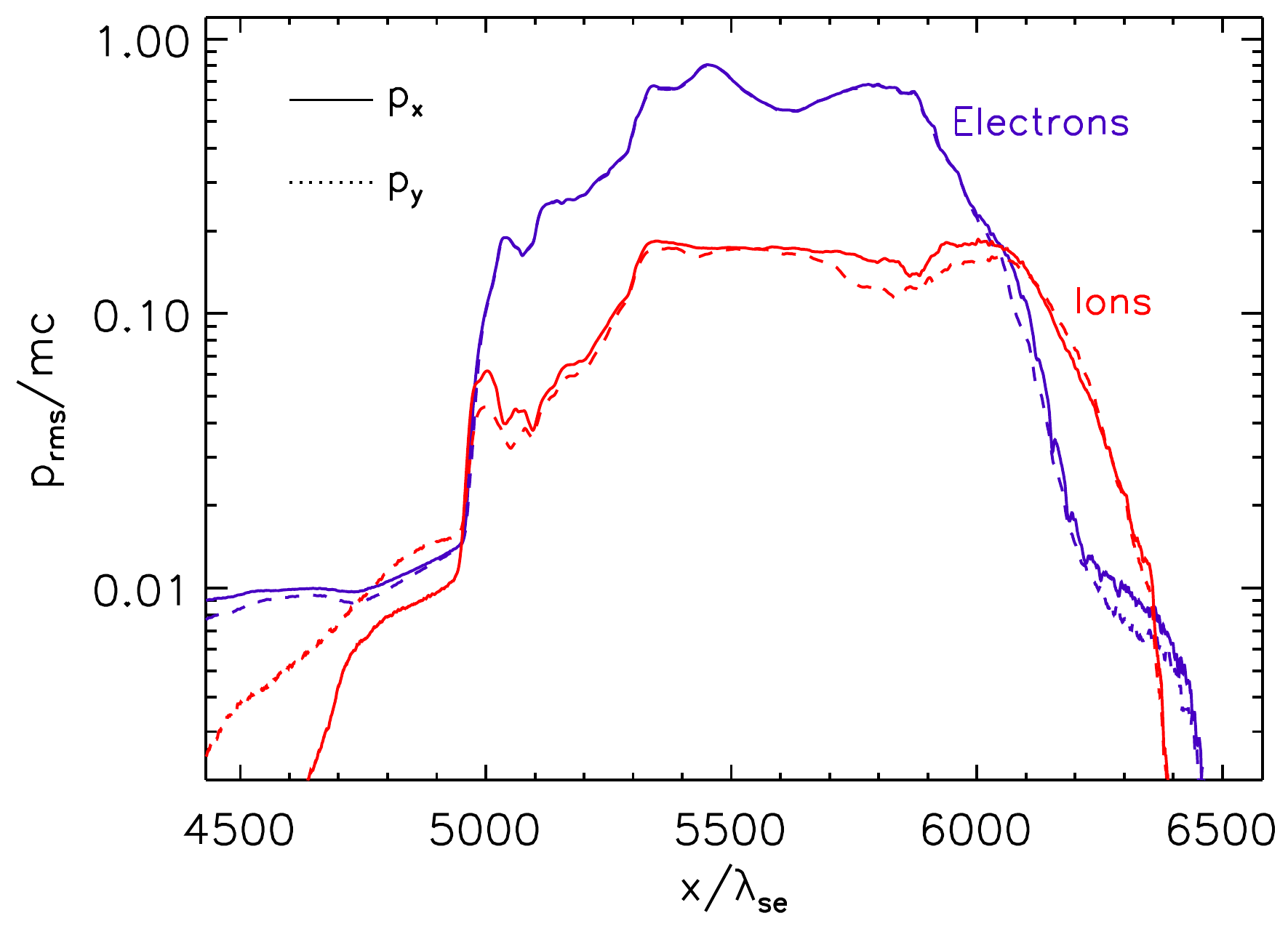}
\caption{RMS values of the momentum components $p_x$ (solid lines) and $p_y$ (dotted lines) of electrons and ions.}
\label{rms_early}
\end{figure}

We follow \citet{amahosh_2012} in the analysis of this first stage of electron heating. The relative velocity of incoming electrons and reflected ions must be larger than the thermal speed of the electrons which leads to the condition
\begin{equation}
M_\mathrm{s}\ge \frac{1+\alpha}{2}\,\sqrt{\frac{m_\mathrm{i}}{m_\mathrm{e}}}\,\sqrt{\frac{T_\mathrm{e}}{T_\mathrm{i}}}\ ,
\label{eq1}
\end{equation}
where $\alpha$ denotes the density ratio of reflected and incoming ions. Whereas the temperature ratio can deviate from unity on account of physics, the dependence on the mass ratio is important for the proper interpretation of simulations that may use a small mass ratio for computational reasons. \citet{amahosh_2012} write this condition in terms of the Alfv\'enic Mach number and the electron plasma beta, which we consider not helpful because the Alfv\'en speed cancels in their expression. The magnetic field may be relevant, however, for the energy gain that electrons may achieve.
If we express the momentum increment as the product of the force provided by the Buneman electric field, $E_\mathrm{B}$, and the Larmor time, $\Omega_\mathrm{e}^{-1}$, then we find
\begin{equation}
\delta p\simeq m_\mathrm{e}\,c\,\frac{E_\mathrm{B}}{B_0}\ .
\label{eq2}
\end{equation}
Assuming isotropization and relating the saturation-level energy density of the electric field to a fraction $0.25\,(m_\mathrm{e}/m_\mathrm{i})^{1/3}$ \citep{1980PhRvL..44.1404I}, one arrives at 
\begin{equation}
\delta p_\mathrm{x,y,z}\simeq \frac{1}{\sqrt{3}}\,\frac{m_\mathrm{e}\,c\,M_\mathrm{A}}{1+\alpha}\,\left(\frac{m_\mathrm{e}}{m_\mathrm{i}}\right)^\frac{2}{3}\simeq 0.9\,m_\mathrm{e}\,c\ ,
\label{eq3}
\end{equation}
where the last, numerical expression is calculated using the parameters of our simulation.
This is evidently much more than the spread of the longitudinal phase-space distribution of electrons, displayed in Figure~\ref{rms_early}, in which the peak intensity of Buneman waves is seen near $x=6100\,\lambda_\mathrm{se}$, where the spread in the $p_x$ component increases to only $\sim 0.1\,m_e\,c$. The distribution of electrons is reasonably close to a Gaussian. The initial acceleration is in x-direction, and the perpendicular magnetic field provides efficient deflection into the y- or z-direction over a distance of $\sim 20\,\lambda_\mathrm{se}$. The ion momentum spread in the upstream regions is dominated by the beam of reflected ions. 

Perhaps estimate of Equation~\ref{eq3} is too optimistic because the electrons lose resonance with the Buneman modes in less than the Larmor time, $\Omega_\mathrm{e}^{-1}$. Note also that the total convertable electron drift energy density \citep{1980PhRvL..44.1404I,amahosh_2012}, when turned into electron heat, gives only
\begin{equation}
\delta p_\mathrm{x,y,z}\simeq \frac{1}{\sqrt{3}}\,\frac{1}{1+\alpha}\,\frac{v_\mathrm{sh}}{c}\,\left(\frac{m_\mathrm{e}}{m_\mathrm{i}}\right)^\frac{1}{6}\simeq 0.1\,m_\mathrm{e}\,c\ ,
\label{eq4}
\end{equation}
which is close to the observed spread of the $p_x$ phase-space component.

For the reverse shock the estimate leading to Equation~\ref{eq3} would also indicate acceleration by $\delta p_x \simeq 0.9\,m_e\,c$, because its Alfv\'enic Mach number does not significantly differ from that of the forward shock. The more conservative estimate based on beam energy conversion (Eq.~\ref{eq4}) suggests $\delta p_x \simeq 0.03\,m_e\,c$, which is 3 times the spread in $p_x$ or $p_y$ that is assumed near $x=4900\,\lambda_\mathrm{se}$ according to Figure~\ref{rms_early}. Note that the wavelength of Buneman modes at the foot of the reverse shock is a small multiple of the grid scale, and so the smoothing and coarse pixelization used to produce Figure~\ref{rv_shock} render the waves invisible in the $E_x$ or density distribution. A Fourier analysis of $E_x$ in the region $\left[4800\,\lambda_\mathrm{se};4900\,\lambda_\mathrm{se}\right]$ reveals a broad peak, corresponding to a parallel mode with $k_\parallel\,\lambda_\mathrm{se}\approx 3\dots 7$. The local electron skin length is $\lambda_\mathrm{se}$, and the resonance condition yields, in analogy to Equation~\ref{eq0}, a range of relative velocities, $v_\mathrm{rel}=0.15\,c\dots 0.3\,c$. Figure~\ref{rv_shock} suggest that there are actually 2 streams of ions that propagate relative to the electrons, one with $0.15\,c$ and the other one with $0.25\,c$. We conclude that the broad bump in the Fourier spectrum indicates the presence of two separately excited modes that we cannot resolve individually.

At the two shocks we can also see a density and magnetic overshoot, i.e., the compression at the shock front with $n_\mathrm{ov}/n_\mathrm{up}\approx 8$ surpasses the expected magnetohydrodynamic compression ratio of $\sim 4$. The rms-amplitude of the magnetic field is about 8.5 times that in the far-upstream region. Such overshoots are common, known for a long time, and result from the flux of returning shock-reflected ions, \citep[e.g.,][]{1983PhFl...26.2742L}. The modest adiabatic heating associated with this excess compression is marginal compared to the heating that the electrons experience at the shock ramp, where an oblique compressive wave mode is evident. We shall discuss these waves together with a full account of particle acceleration for the final state of the simulation, at $t\,\Omega_i=20$.

\subsection{Late-stage Evolution}\label{late_stage}
\begin{figure}[htb!]
\centering
\includegraphics[width=\linewidth]{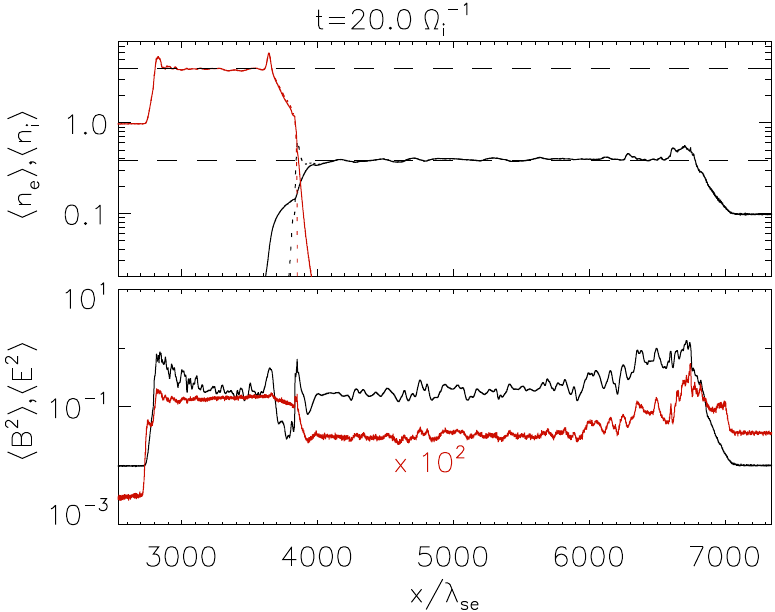}
\caption{Structure of the collision region at the end of the simulation at time $t=20\,\Omega_\mathrm{i}^{-1}$. Shown are in the top panel the profiles of the average particle-number density, compared with horizontal {long-}dashed lines marking the expected compression level of $n_{R,d}/n_{R,u}=3.86$ for the forward shock (lower line) and $n_{L,d}/n_{L,u}=4.02$ for the reverse shock (upper line). In the bottom panel we display the average magnetic energy density in black and the electric energy density in red, the latter scaled with a factor 100.}\label{plot_late_stage}
\end{figure}

We let our perpendicular shock simulation run for $20\,\Omega_\mathrm{i}^{-1}$. Figure~\ref{plot_late_stage} presents evidence of the very clear separation of the three discontinuities. The shocks are located at around $x\approx6750\,\lambda_\mathrm{se}$ (forward shock) and $x\approx2800\,\lambda_\mathrm{se}$ (reverse shock). The shock speeds measured in the simulation frame are consistent with the expected values of $v_\mathrm{sh,R}=0.036\,c$ for the forward shock and $v_\mathrm{sh,L}=-0.093\,c$ for the reverse shock. The CD has reached a position at $x\approx3850\,\lambda_\mathrm{se}$, consistent with its predicted speed of $v_{\mathrm{CD}}=-0.06\,c$.

To be noted from Figure~\ref{plot_late_stage} is that downstream of the overshoot regions the compression ratio at the shocks is very well in agreement with the hydrodynamical jump conditions for non-relativistic gas with $\Gamma=5/3$, which predict $n_{R,d}/n_{R,u}=3.86$ and $n_{L,d}/n_{L,u}=4.02$ in the simulation frame for the forward and reverse shocks, respectively (upper and lower dashed lines in the top panel of Figure \ref{plot_late_stage}). We can therefore expect that the system has reached a statistical equilibrium for the bulk of the plasma, and a detailed analysis of the shock structure, shock reformation, and particle acceleration can be conducted.
We shall discuss the two shocks in turn.

\subsubsection{Structure of the Forward Shock}\label{forward_shock}
\begin{figure}[htb!]
\centering
\includegraphics[width=\linewidth]{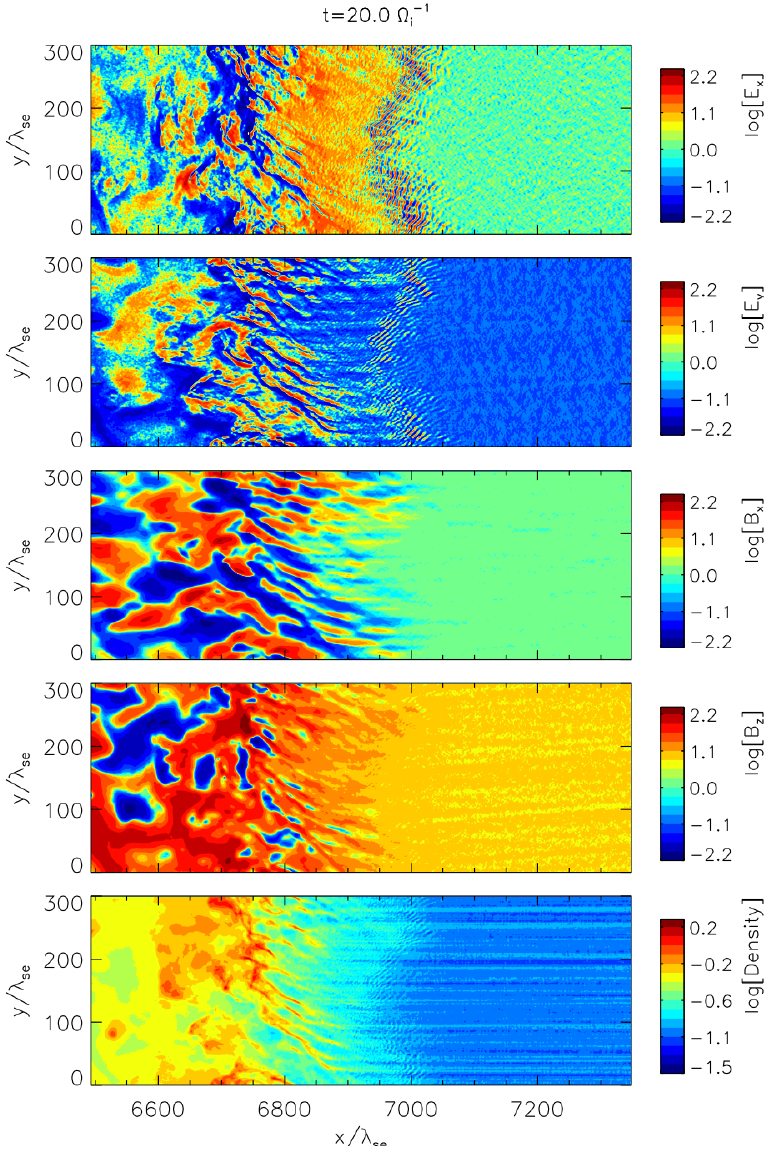}
\caption{Structure of the forward shock at the end of the simulation at time $t=20\,\Omega_\mathrm{i}^{-1}$. Shown are from top to bottom the field components $E_x$, $E_y$, $B_x$, and $B_z$, followed by the electron density. As in earlier plots, we use a sign-preserving logarithmic scale for the field amplitudes.}
\label{plot_forward_shock}
\end{figure}

\begin{figure}[htb!]
\centering
\includegraphics[width=\linewidth]{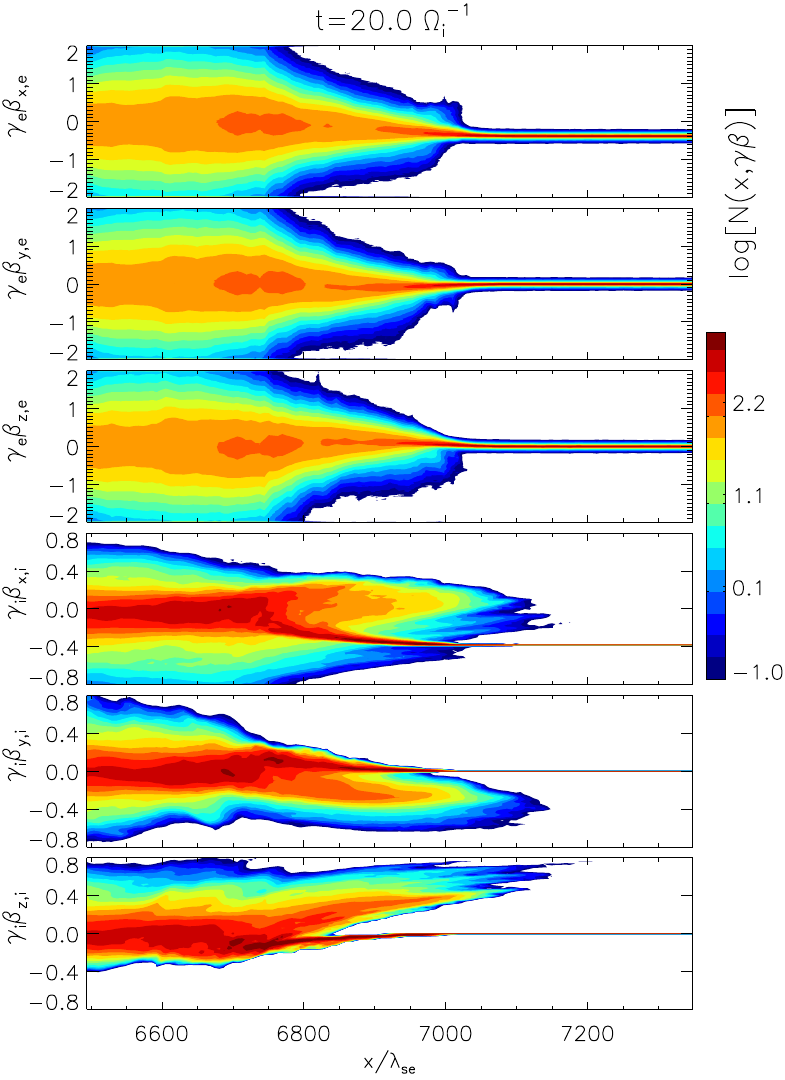}
\caption{Phase-space distribution near the forward shock at the end of the simulation.}\label{plot_forward_shock_ps}
\end{figure}

Figures~\ref{plot_forward_shock} and \ref{plot_forward_shock_ps} present the structure of the forward shock at the end of the simulation at time $t=20\,\Omega_\mathrm{i}^{-1}$. Displayed in Figure~\ref{plot_forward_shock} are the density distribution of the electrons and the amplitudes of two components of the electric field and the magnetic field. The distribution of the ions (not shown) closely follows that of the electrons, except where the latter is modulated by the Buneman instability. Figure~\ref{plot_forward_shock_ps} presents the phase-space distribution of electrons and ions.

\begin{figure*}[htb!]
\centering
\includegraphics[width=0.33\linewidth]{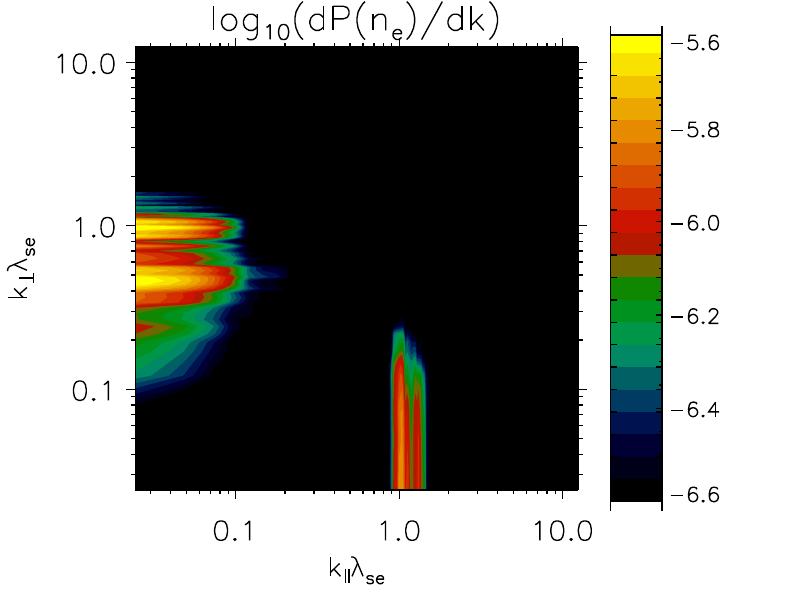}
\includegraphics[width=0.33\linewidth]{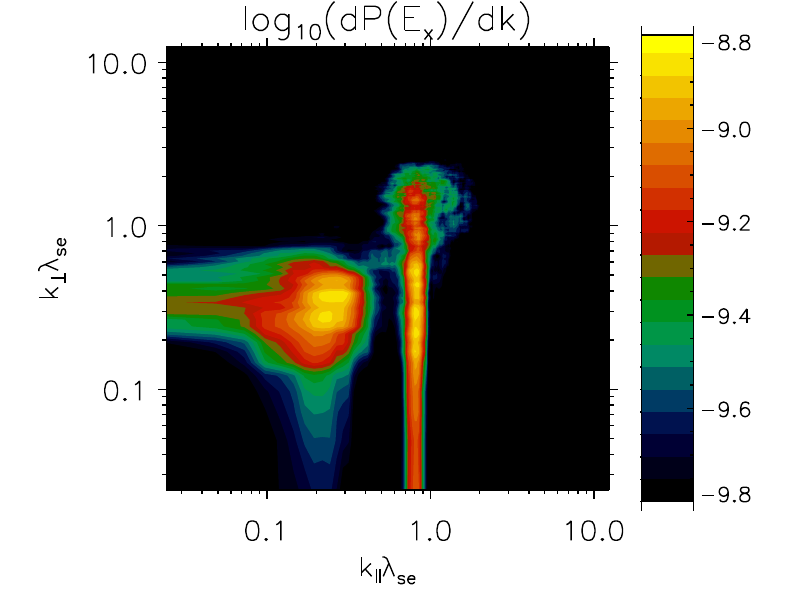}
\includegraphics[width=0.33\linewidth]{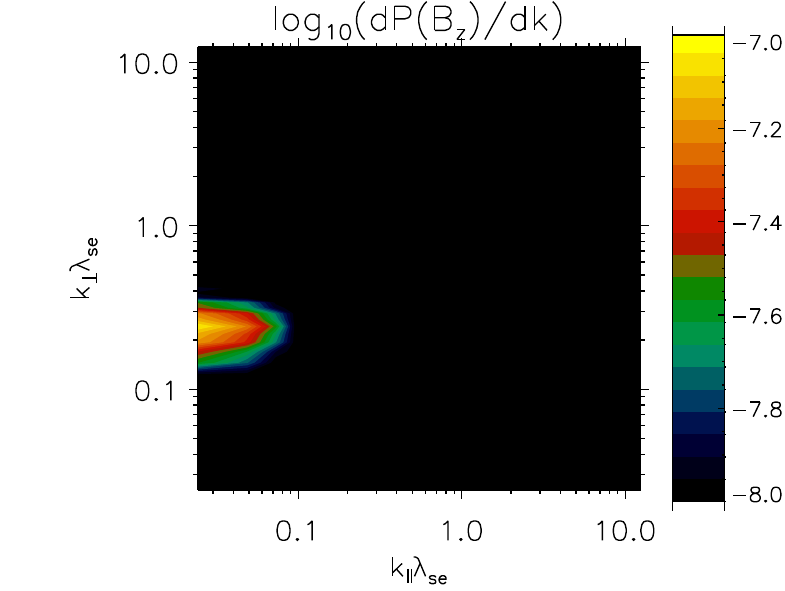}
\caption{Fourier power spectra taken in the region $7100\,\lambda_\mathrm{se}$ -- $7360\,\lambda_\mathrm{se}$. Shown in two-dimensional reduced wavevector space $\left(Z_{\parallel},Z_{\perp}\right)=\left(k_{\parallel}\,\lambda_\mathrm{se},k_{\perp}\,\lambda_\mathrm{se}\right)$ are spectra of the electron density (left), the electric-field component $E_x$ (middle), and the magnetic-field component $B_z$ (right).}\label{fourier}
\end{figure*}

While strong Buneman-type turbulence can be seen directly in the $E_x$ and density distributions around $x=6950\,\lambda_\mathrm{se}$ (see Fig.~\ref{plot_forward_shock}), a weaker electrostatic mode is seen throughout the precursor, as well as filamentation. The complexity of the turbulence in the precursor region is demonstrated in Figure~\ref{fourier}, where we show two-dimensional power spectra of the electron density, the electric-field component $E_x$, and the magnetic-field component $B_z$. The main Buneman modes have $k_\parallel\,\lambda_\mathrm{se}\simeq 0.8$
(Fig.~\ref{plot_forward_shock}b), as predicted in Equation~\ref{eq0}, and can be slightly oblique \citep{1974PhFl...17..428L}.  The associated density fluctuations are seen only at slightly larger $k_\parallel$ and for small $k_\perp$ (Fig.~\ref{plot_forward_shock}a). The second mode in the $E_x$ power spectrum, located at $\left(k_{\parallel}\,\lambda_\mathrm{se},k_{\perp}\,\lambda_\mathrm{se}\right)\approx (0.25,0.4)$, has no counterpart in the density spectrum, but it is weakly seen in the spectra of other electric or magnetic field components as well.

The filamentation mode in the precursor is seen in $B_z$ (and also in $B_y$) at $k_{\perp}\,\lambda_\mathrm{se}\approx 0.25$ (Fig.~\ref{plot_forward_shock}c), {equivalent} to a wavelength $\lambda\approx 8\,\lambda_\mathrm{se,local}$, which corresponds to the separation of current filaments. In the density spectra, however, we see two {other} peaks at $\lambda\approx 2\,\lambda_\mathrm{se,local}$ and $\lambda\approx 4\,\lambda_\mathrm{se,local}$ that are the dominant first two harmonics needed to represent the size of the current filaments, which we estimate as $\sim 3\,\lambda_\mathrm{se,local}$.

\begin{figure}[htb!]
\centering
\includegraphics[width=\linewidth]{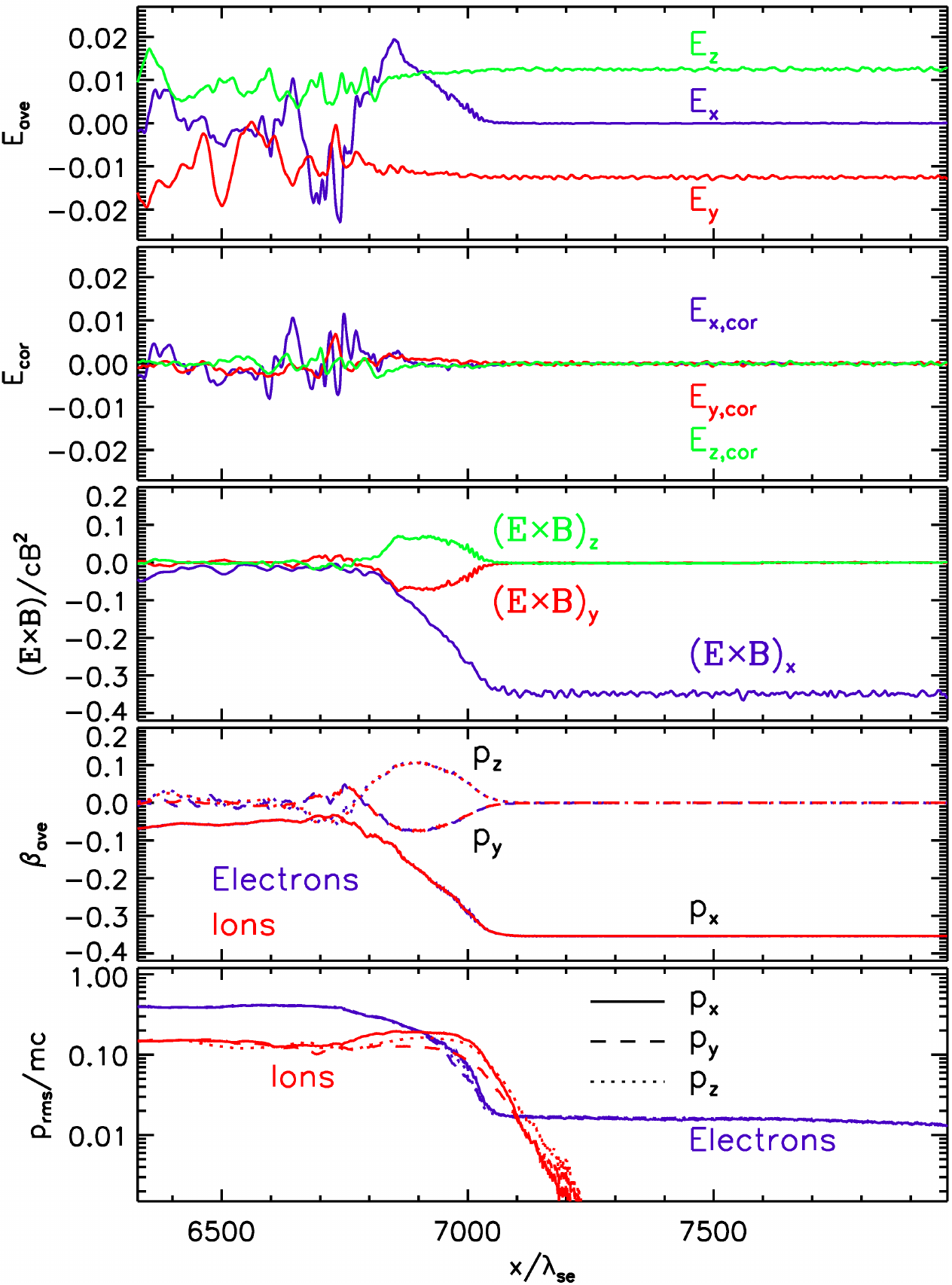}
\caption{Moments of the phase-space distribution of electrons and ions in comparison to the electric-field components {in the simulation frame and in the local flow frame given by $\mathbf{\beta}_\mathrm{ave}$,} and the components of $\mathbf{E}\times\mathbf{B}$ drift. All quantities are averaged over the $y$-coordinate. For ease of comparison with the drift speed we plot the average motion as velocity $\beta=p/mc(1+p^2)$.}\label{moments}
\end{figure}

Overall, the amplitude of waves in the precursor is low and does not increase the field energy density. It is sufficient, however, to keep the electrons warm. Figure~\ref{moments} displays as function of the $x$-coordinate the $y$ averages of the moments of the electron and ion phase-space distributions, the electric field, and the $\mathbf{E}\times\mathbf{B}$ drift. The normalized electron momentum spread is nearly constant at $p_\mathrm{rms}\simeq 0.02\,mc$ throughout the precursor, about a factor 10 larger than at injection. We cannot exclude that some of the heating is related to turbulence caused by penetrating or reflected particles from the initial plasma collision. 

Further heating can occur at the foot of the shock. {Note that} the Alfv\'enic Mach number is too large, or the mass ratio between ions and electron too small, for Whistler waves to be generated in the foot region \citep{2006JGRA..111.6104M}, that in low-Mach-number simulations were observed to energize electrons \citep{riqspit_2011}.  

In the shock ramp and further downstream the electric field shows large fluctuations. Similar variations are seen in the magnetic field, but already after smoothing over scales smaller than the electron Larmor radius the $\mathbf{E}\times\mathbf{B}$ drift speed has a rather smooth distribution. The shock ramp is clearly delineated by the gradient in $(E\times B)_x$ between $x\simeq 6800\,\lambda_\mathrm{se}$ and $x\simeq 7000\,\lambda_\mathrm{se}$, as shown in Figure~\ref{moments}. Over the entire shock ramp $\mathbf{E}\times\mathbf{B}$ drift with $v/c\lesssim 0.1$ should occur in positive $z$ and negative $y$ direction.

A gradient drift is also expected and should lead to shock-drift acceleration \citep[henceforth called SDA][]{1989JGR....9415089K,sda_1989}. It is important to note, though, that our strictly perpendicular shocks are superluminal, hence a de-Hoffman-Teller frame does not exist, {and injection into SDA should be suppressed} \citep{2001PASA...18..361B}. 

In the literature one usually finds the gradient drift discussed in the context of a smooth gradient in the large-scale magnetic field. Written in the downstream frame, i.e. the CD frame, the global gradient of the magnetic field near a nonrelativistic shock front causes a particle of mass $m$ and charge $q$ to drift with velocity \citep[e.g.][]{guo_sironi_2014}
\begin{align}
v_\mathrm{gd} &=\frac{p_\perp^2}{2\,m\,q\,B}\,\frac{\mathbf{B}\times\mathbf{\nabla}B}{B^2}
\nonumber \\
 &=\frac{\mathrm{sgn}(q)\,p_\perp^2}{2\,m\,m_i\,\Omega_i}\,\frac{\mathbf{B}\times\mathbf{\nabla}B}{B^2}  \ ,
 \label{eq5}
\end{align}
where $p_\perp$ is the momentum perpendicular to the magnetic field. The first point to notice is the dependence on the charge sign, $\mathrm{sgn}(q)$, which mandates that electrons and ions drift in opposite directions. The second fraction on the right-hand side of Equation~\ref{eq5} must on average be similar to the inverse thickness of the shock ramp, $\sim \Omega_i/v_\mathrm{sh}$. Noting further that momentum conservation requires that on average $p_\perp^2 \simeq (3/8)\,m^2\,v_\mathrm{sh}^2$, Equation~\ref{eq5} can be simplified to
\begin{equation}
\vert v_\mathrm{gd}\vert \lesssim \frac{3}{16}\,\frac{m}{m_i}\,v_\mathrm{sh}\simeq 0.075\,c\,\frac{m}{m_i}\ ,
\label{eq6}
\end{equation}
or about $0.05\,c$ projected along the $y$ and $z$ axes, indicating that whereas the ions may drift with a velocity not much smaller than that of $\mathbf{E}\times\mathbf{B}$ drift, electrons would be considerably slower. 

{Standard descriptions of SDA, including the lack of injection into it at superluminal shocks, rely on the concept of a smooth shock transition, in which the magnetic-field gradient is a clearly discernible feature. One consequence is that gradient drift should exist, but here it doesn't. Turbulence in the ramp destroys this picture, and the true electromagnetic environment likely deviates from the simple concept, as demonstrated, e.g., on page 6 of \citet{amahosh_2012} and attributed there to time dependence arising from shock reformation.}
The large fluctuations in the electric and magnetic field in the shock ramp play havoc with the gradient drift, because small- and medium-scale structures provide by far dominant contribution to $\mathbf{\nabla}\,B$. We could not find an averaging scheme that permits extracting from the simulation data a gradient drift corresponding to that calculated for the globally expected compression of the large-scale magnetic field.

The observed mean velocity, $\mathbf{\beta}_\mathrm{ave}=\mathbf{v}_\mathrm{ave}/c$, follows closely that expected from $\mathbf{E}\times\mathbf{B}$ drift. There is no discernible difference in the flow of electrons and ions, indicating the absence of a significant gradient drift. To be noted is that $\mathbf{E}\times\mathbf{B}$ drift is not a property of the reflected particles alone. The average particle velocity corresponds to the drift speed, implying that reflected particles have transverse speeds a factor of a few higher than the drift speed.

{We can transform the electric field into the local flow frame given by $\mathbf{\beta}_\mathrm{ave}$. The resulting field in this guiding-center frame, $\mathbf{E}_\mathrm{cor}$, is displayed in the second panel of Figure~\ref{moments}, and to be noted is the low amplitude throughout the shock ramp. 
We emphasize the important lesson that the existence of an electric-field component parallel in the drift direction of particles does not imply the existence of a significant electric field in the drift frame.

Plasma crosses the ramp in a time $t_\mathrm{rc}\simeq \Omega_i^{-1}$. For individual ions the separation of motion into drift and gyration is thus questionable, whereas for electrons a guiding-center approximation should work well. Averaged over their Larmor motion, electrons see only very weak electric fields, and hence are not significantly accelerated beyond the adiabatic compression at the shock unless their Larmor radius is of the same order or larger than the ramp thickness on account of pre-acceleration at the foot or further upstream. Ions generally fulfil that condition, and so they may see a coherent electric field along their trajectory across the ramp and be accelerated.

The ion distribution in the ramp is composed of incoming, reflected, and returning reflected particles whose superposition may provide a reasonable sampling of gyrophase, thus retaining the guiding-center approximation crude but still reasonable for $\mathbf{E}\times\mathbf{B}$ drift, hence the agreement between expected and observed drift speed. The significant increase in $p_\mathrm{rms}$ of the ions at the foot of the shock between $x=7000\,\lambda_\mathrm{se}$ and $x\simeq 7100\,\lambda_\mathrm{se}$ does not indicate heating but reflects the momentum offset between incoming and reflected particles. A similar effect is expected for electrons, but would not be as easily identifiable on account of their small Larmor radius. }

\begin{figure}[htb!]
\centering
\includegraphics[width=\linewidth]{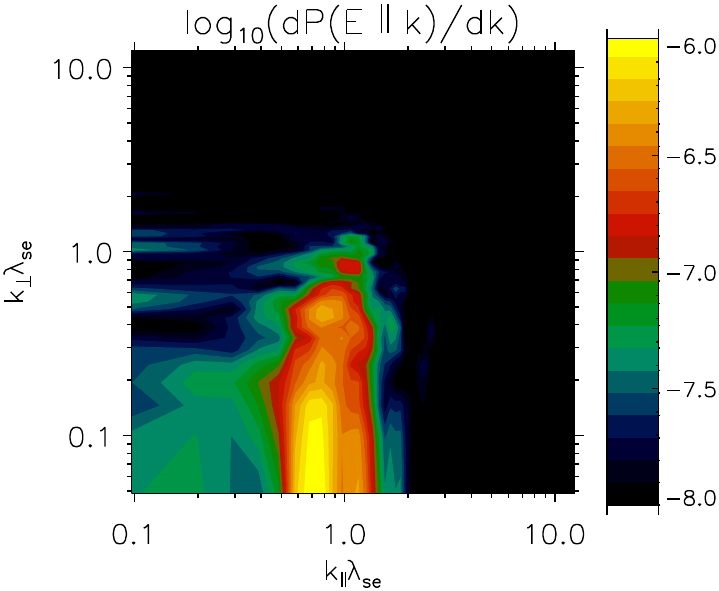}
\caption{Fourier power spectrum of the electric field parallel to the wave vector, $\mathbf{e}_k\cdot\mathbf{E}$, in the region $6970\,\lambda_\mathrm{se}< x <7050\,\lambda_\mathrm{se}$ and $y\le 125\,\lambda_\mathrm{se}$.}\label{epar}
\end{figure}
In section~\ref{distributions} we shall discuss in detail the resulting particle spectra in the downstream region. Here, we only note that the normalized moment spread, $p_\mathrm{rms}/mc$, of electrons in the far downstream region evolves to 
about four times that of the ions. The ratio of electron temperature to that of ions, if thus defined, is 
then
\begin{equation}
\frac{T_e}{T_i}\simeq 0.28\,,
\label{eq7}
\end{equation}
much larger than the $1/50$ expected for purely magnetic isotropization but still short of equilibration. Observationally one finds, with few exceptions, a correlation $T_e/T_i \propto v_\mathrm{sh}^{-2}$ \citep[for a review][]{2013SSRv..178..633G}. A conceptional explanation for this empirical relation lies in electron heating through turbulence generated by reflected ions whose number depends on the shock speed. It is not easy to relate the temperature relation found in our simulation to those inferred for real SNRs, because we use a reduced mass ratio and the empirical relation does not distinguish between the different efficiency in ion reflection of perpendicular and parallel shocks. 

\begin{figure}[htb!]
\centering
\includegraphics[width=\linewidth]{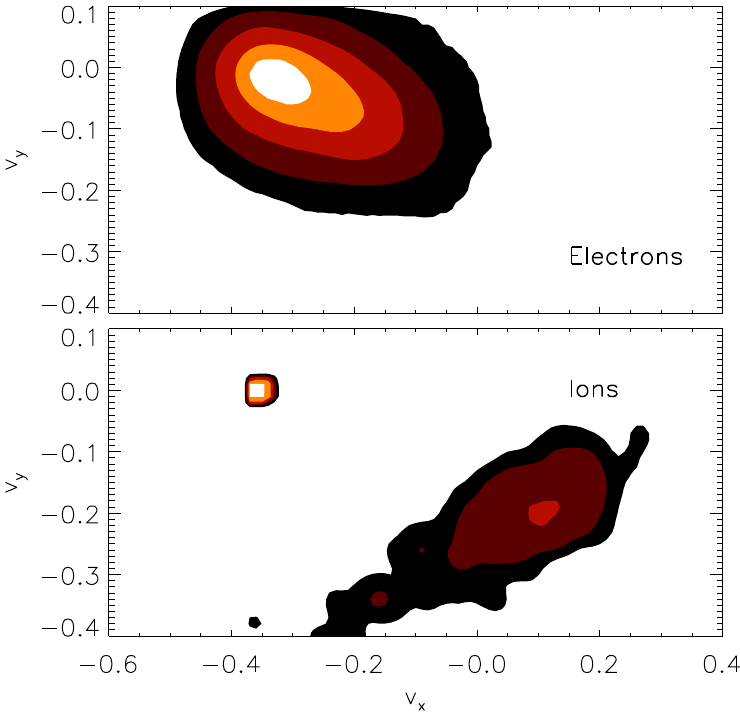}
\caption{Phase-space distribution in $v_x$-$v_y$ of electrons (Top panel) and ions (Bottom) in the section of the foot region from which we extracted the Fourier spectrum shown in Fig.~\ref{epar}. The scale is logarithmic with maximum at 40\% of the peak density. The dynamic range is 33000 for electrons and 230000 for ions.}\label{ps-foot}
\end{figure}
The Buneman modes excited at the shock foot may give rise to shock-surfing acceleration of electrons \citep{leroy1981,amahosh_2009,amahosh_2012}. Despite the electron heating in the far-upstream region, the high sonic Mach number of the shock exceeds the threshold for the growth of Buneman modes (see Eq.~\ref{eq1}) as first stage of electron acceleration. Visual inspection of Figure ~\ref{plot_forward_shock} reveals non-planarity of the foot region, i.e. rippling {(see Sec.~\ref{reverse_shock})}. 
 We {therefore} select a small region ($6970\,\lambda_\mathrm{se}< x <7050\,\lambda_\mathrm{se}$ and $y\le 125\,\lambda_\mathrm{se}$) for further analysis, that appears to harbor a planar sheet of strong electrostatic fluctuations. Figure~\ref{epar} displays a Fourier spectrum of the electric field parallel to the wave vector, $\mathbf{e}_k\cdot\mathbf{E}$. We observe a strong signal at $k_\parallel\,\lambda_\mathrm{se}\simeq 0.7$, as is seen already during the early evolution and is in fact expected (see Eq.~\ref{eq0}). This is clearly a Buneman mode and not an ion-acoustic wave which would have a much larger wavenumber, $k$.  \citet{amahosh_2012} conducted 2-D PIC simulations of non-relativistic shocks with Alfv\'enic Mach number and ion/electron mass ratio similar to those in our work. They find an electrostatic mode with $k_\perp/k_\parallel \simeq 0.5$ whereas in our simulation $k_\perp/k_\parallel\simeq 0.15$ (and slightly larger than that for regions shifted along the transverse coordinate, $y$). The linear analysis of \citet{2009PhPl...16j2901A} suggests that oblique modes should be strong for electron-ion drift much faster than the electron thermal velocity, whereas the parallel mode should dominate right above the threshold of the instability. Indeed, a Fourier spectrum taken further upstream (for $x >7050\,\lambda_\mathrm{se}$) shows a prominent oblique mode, albeit with an intensity a factor of 10 lower than that observed right at the foot. 

It is instructive to compare the $v_x-v_y$ ion phase-space density distribution in our simulation data to that of \citet[see their Fig.2a]{amahosh_2012} and \citet{2013PhRvL.111u5003M}. In the region where Buneman modes are strong they find reflected ions with substantial $v_y$ on the order of the shock speed. Figure~\ref{ps-foot} displays the phase-space distribution in our simulation, and it is evident that the drift speed between electrons and ions is oriented in various directions. The density peak in the distribution of reflected ions corresponds to approximately $v_y/v_x\simeq 0.3$, and this is in fact the only drift component that is also seen a bit further upstream, indicating that an instability driven by this component was already operating in those regions and thus may have reached a large amplitude. 

The main difference between our simulation setup and that of \citet{amahosh_2012} and \citet{2013PhRvL.111u5003M} is that they assume the large-scale magnetic field strictly out of plane, and that their electron plasma beta is close to unity, whereas for us it is on the order $1/25$ even after electron heating in the far-upstream region. Accounting for magnetic-field orientation yields a factor $\sqrt{2}$, as only $1/\sqrt{2}\simeq 0.71$ of the large-scale magnetic deflection projects on the $y$ axis, and we still find both the direction of reflected ions and the wave vector of Buneman modes aligned closer to the shock normal than is the case in the simulation of \citet{amahosh_2012}. As it is unclear how the plasma beta could cause this discrepancy, its origin remains unclear. In general, the non-Maxwellian form of particle distribution functions and their rapid change in the foot and ramp regions renders difficult a linear analysis of wave types and instabilities.

\begin{figure*}[htb!]
\centering
\includegraphics[width=0.274\linewidth]{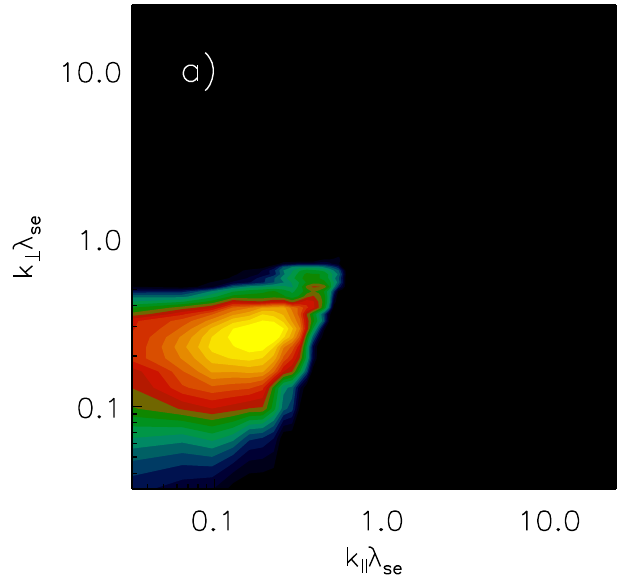}
\includegraphics[width=0.32\linewidth]{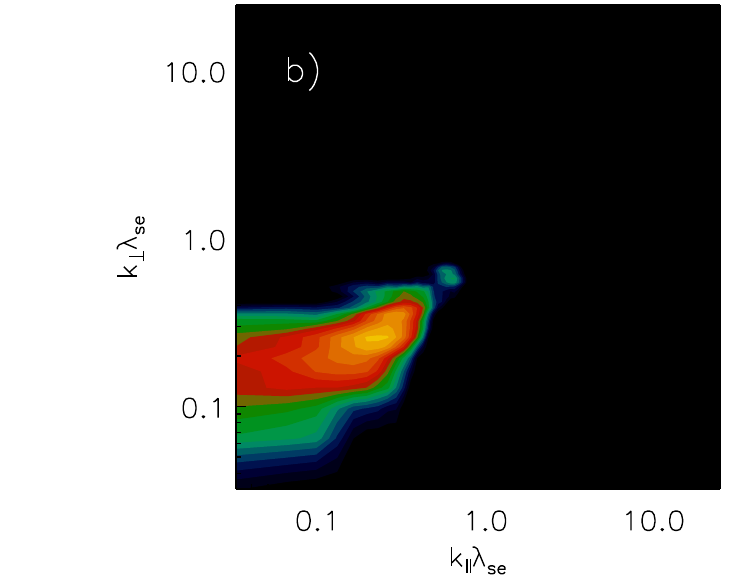}
\includegraphics[width=0.39\linewidth]{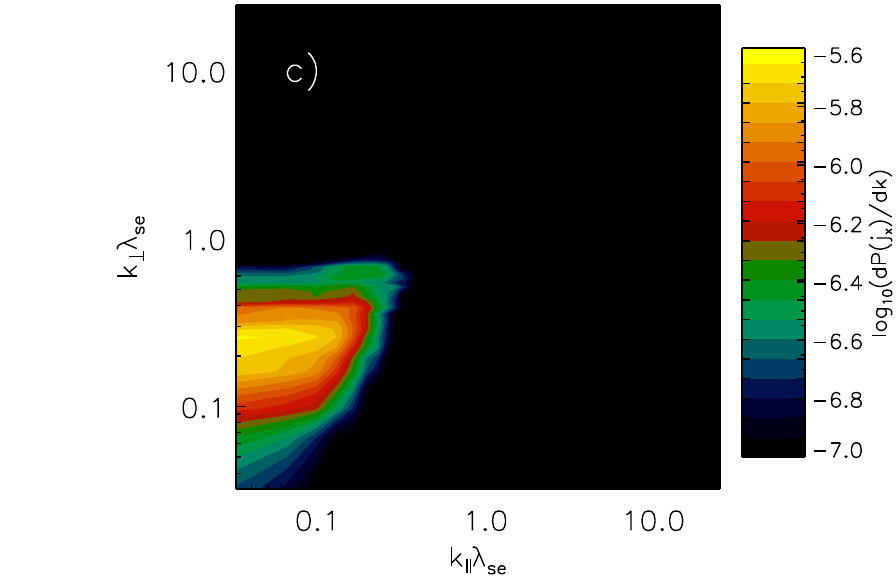}
\caption{Fourier spectra of the current density in $x$ direction, $\mathbf{j}_x$, in the ramp region $6760\,\lambda_\mathrm{se}< x <6960\,\lambda_\mathrm{se}$. Panel a) on the left displays the  spectrum for nominal coordinates $x$ and $y$, whereas for panels b) and c) the $y$ coordinates were shifted before computing the Fourier spectrum. For panel b) we compensated the average velocity of all ions (see Figure~\ref{moments}), whereas for panel c) on the right we compensated only the average motion of incoming ions.}\label{four_current}
\end{figure*}

In the ramp region a nearly perpendicular electromagnetic mode is evident in the field components shown in Figure~\ref{plot_forward_shock}. Similar filaments were observed in the foot region in the simulation of \citet{kato_2010} and identified as current filaments resulting from the Weibel-type instabilities that were shown to mediate high-speed nonrelativistic unmagnetized and weakly mangetized parallel shocks \citep{kato2008,niemiec_2012}. We conducted a Fourier analysis and indeed found strong evidence of filamentation in $\mathbf{j}_x$ and $\mathbf{j}_z$ (out of plane) that is carried largely by incoming particles. Figure~\ref{four_current} compares Fourier spectra of $\mathbf{j}_x$ that were computed for nominal coordinates $x$ and $y$  with those calculated after shifting the $y$ coordinates to compensate for the curved large-scale trajectories of plasma. We find a nearly perpendicular mode, indicating straight current filaments, only when accounting for the average velocity of incoming ions alone (Fig.~\ref{four_current}c). The wavelength is about two local ion skin lengths, $\lambda\approx 2\,\lambda_\mathrm{si,loc}$, the main uncertainty arising from averaging the local skin length over the density gradient in the shock ramp. Equally strong perpendicular fluctuations are seen in $\mathbf{j}_z$, extending in wavelength to $\lambda\approx 4\,\lambda_\mathrm{si,loc}$. The Fourier signal in $\mathbf{j}_y$ shows a peak at similar scales, but is weaker. If we separately consider the current density carried by electrons and that of ions, we find large Fourier amplitudes for perpendicular scales larger than $2\,\lambda_\mathrm{si,loc}$ that seem to compensate each other, suggesting a density structure that is more complex than that seen in $\mathbf{j}_x$. 

We do not see any evidence for spontaneous turbulent reconnection in the ramp that has been recently reported by \citet{2015Sci...347..974M}. They noted a chain of magnetic islands coinciding with spikes in electron density resembling structures seen in dedicated simulations of turbulent reconnection. There, the neutral lines appear to break up into contracting magnetic islands, in which trapped electrons are accelerated through a Fermi-like process \citep{2006Natur.443..553D}. The filaments in our simulations are only weakly modulated on a scale similar to that of the Buneman waves in the foot, that the electrons passed through on their way up the ramp. Further modulation arises from filament mergers, but the amplitudes are moderate and they also do not align into a chain of magnetic islands.

\subsubsection{Structure of the Reverse Shock}\label{reverse_shock}

\begin{figure}[htb!]
\centering
\includegraphics[width=\linewidth]{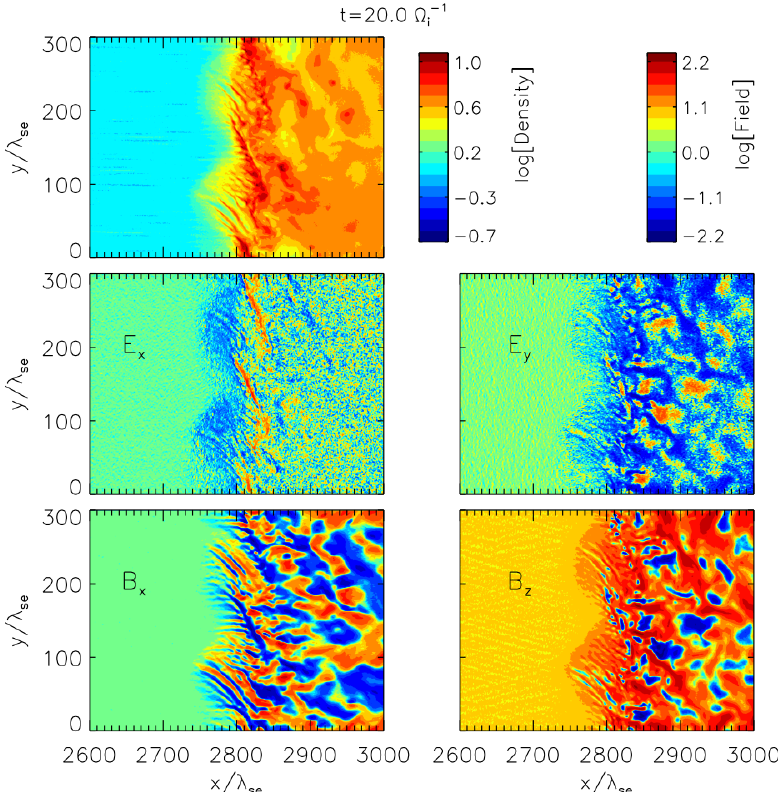}
\caption{Structure of the reverse shock at the end of the simulation. Shown are with proper aspect ratio the electron density in the top left and the field components $E_x$, $E_y$, $B_x$, and $B_z$. As in earlier plots, we use a common sign-preserving logarithmic scale for the field amplitudes, displayed in the upper right corner.}\label{plot_reverse_shock_ps}
\end{figure}

Figure~\ref{plot_reverse_shock_ps} presents the structure of the reverse shock at the end of the simulation at time $t=20\,\Omega_\mathrm{i}^{-1}$. Displayed are the density of electrons and the amplitudes of four components of the electromagnetic field. In terms of shock parameters, the main difference to the forward shock is a significantly lower sonic Mach number, $M_\mathrm{s}=252$ instead of $M_\mathrm{s}=755$. Both the shock speed and the skin lengths are smaller by about a factor 3 compared to their values at the forward shock, and so the reverse shock extends over fewer grid points and is not as well resolved by the simulation.

{To be noted from Figure~\ref{plot_reverse_shock_ps} is that the shock surface is highly perturbed, and these shock-front fluctuations can be identified as ripples.}  
{
They occur on spatial scales of about 
$150\,\lambda_\mathrm{se}\approx 20\,\lambda_\mathrm{si}$ and are visible through correlated fluctuations in density and all electromagnetic field components. Time analysis reveals that the ripples propagate in $+y$-direction with an average speed of $0.08\,c\approx 18\,v_\mathrm{A, L}$. Their structure is highly dynamic on timescales as short as $0.1\,\Omega_i^{-1}$ and their shape changes, typically toward an asymmetric wave pattern. 
The amplitude of rippling, i.e. the displacement in $x$, varies between $2,\lambda_\mathrm{si}$ and $5,\lambda_\mathrm{si}$. The ripples emerge with small amplitude at the same time as the shock itself ($t\Omega_i=4-5$), and they may initially be influenced by the curved structure of the CD resulting from an incoherent development of filamentation instabilities on both sides of the discontinuity (see Fig.~\ref{plot_double_shock}). The subsequent evolution is highly variable in amplitude and wavelength ($100-300\,\lambda_\mathrm{se}$). The specific form of the ripples shown in Figure~\ref{plot_reverse_shock_ps} is acquired only during the late-stage evolution at $t\Omega_i\approx 16$.}

{Shock rippling is known from numerical studies of quasi-perpendicular supercritical low-Mach number shocks performed in two or three dimensions with transverse size large enough to contain several ion skin lengths \citep[see, e.g.,][]{winske_quest88,lowe_burgess2003,thomas1989,burgess-scholer_2007,umeda2009,umeda2010,umeda2014}. The ion temperature anisotropy arising from ion reflection at the shock can drive the Alfv\'en ion cyclotron (AIC) or the mirror instability in the shock ramp, and the resulting unstable modes have wavelengths of {\it a few} ion skin depths and propagate along the regular magnetic field, significantly contributing to ion isotropization and thermalization at the shock and downstream. Studies of such wave structures with 2D simulations require the in-plane configuration of the magnetic field. 
For out-of-plane field configurations, that suppress parallel-propagating waves, \citet{burgess-scholer_2007} demonstrated the existence of another shock-front instability that produces fluctuations on a spatial scale commensurate with the gyroradius of shock-reflected ions, that propagate along the shock surface with the speed and direction of the ions gyrating in the shock foot. They showed that the instability requires sufficiently high Alfv\'enic Mach numbers, $M_\mathrm{A}$, but still in the range considered here as the low-Mach number regime, or low plasma beta, and can also occur for magnetic-field orientations $\phi\neq 90^{\rm{o}}$.
Although our simulation with $\phi=45^{\rm{o}}$ allows both types of instabilities to develop in the system, the characteristics of the observed ripples agree with the scenario of \citet{burgess-scholer_2007}, i.e. a modulation of shock-reflected ions along the shock surface. The wavelength of the ripples, $150\,\lambda_\mathrm{se}$, is approximately equal to the gyroradius of reflected ions in the upstream magnetic field projected on the $y-$axis. Their propagation speed and direction is also in line with the reflected-ion speed projected onto the shock surface, for which we find $0.07-0.09\,c$ in an analysis of the ion phase-space distribution in the shock foot.
Modes driven by ion-temperature anisotropy are not excited because plasma isotropization at the shock is effected by Weibel-like filamentation instabilities, and electrons are already relativistically hot through interactions with Buneman waves at the shock foot. 

We note that the rippled shock structure can at some time instances also be observed at the forward shock (e.g., Fig.~\ref{plot_forward_shock}). The amplitude of the ripples is similar to that in the reverse shock in terms of the local plasma skin depth. Shock surface fluctuations now move in the $-y$-direction, consistent with the opposite sense of rotation of reflected ion rotation due to the opposite plasma inflow velocity \citep{burgess-scholer_2007}. We thus conclude, that the mechanism of ripple formation is the same as at the reverse shock. 
However, the gyroradius of ions reflected at the forward shock is in the range of $120-200\,\lambda_\mathrm{se,R}=380-630\,\lambda_\mathrm{se}$, larger than the transverse size of our computational box. The development of the ripples at the forward shock is therefore influenced by our boundary conditions. 

Shock-front ripples have not yet been reported in multi-dimensional studies of high-Mach-number perpendicular shocks. Two-dimensional studies in out-of-plane magnetic-field geometry \citep{amahosh_2012,2013PhRvL.111u5003M} suppress AIC waves and typically use computational boxes with transverse size of $5-6\,\lambda_\mathrm{si}$ that cannot contain ion-gyroscale modulations.
Ion-scale ripple structures have not been identified in the large-scale ($L_y\sim 36.5\,\lambda_\mathrm{si}$) 2D simulations of \citet{kato_2010} with in-plane magnetic-field configuration, although a visual similarity of the magnetic-field and density patterns at and behind the overshoot to the structures reported in \citet{winske_quest88} was noted. Results of our simulation with magnetic field inclined at $\phi=45^{\rm{o}}$ to the simulation plane are thus in agreement with these earlier studies, and in addition they suggest that rippling at the gyroscale of ions will significantly contribute to shock-front nonstationarity even in three-dimensional simulations. 

As consequence of the rippled structure the reflection rate of the ions is enhanced in some locations along the shock front, which is clearly visible through the modulated extension of the filaments in the foot. These regions then provide stronger Buneman instability that should lead to a more efficient localized electron heating and acceleration (see \citet{umeda2009} for similar effects in low-Mach-number shocks). Due to the poorly resolved foot structure at the reverse shock, a detailed analysis of this possible connection between shock ripples and electron energization must be deferred to future work.}

\begin{figure}[htb!]
\centering
\includegraphics[width=\linewidth]{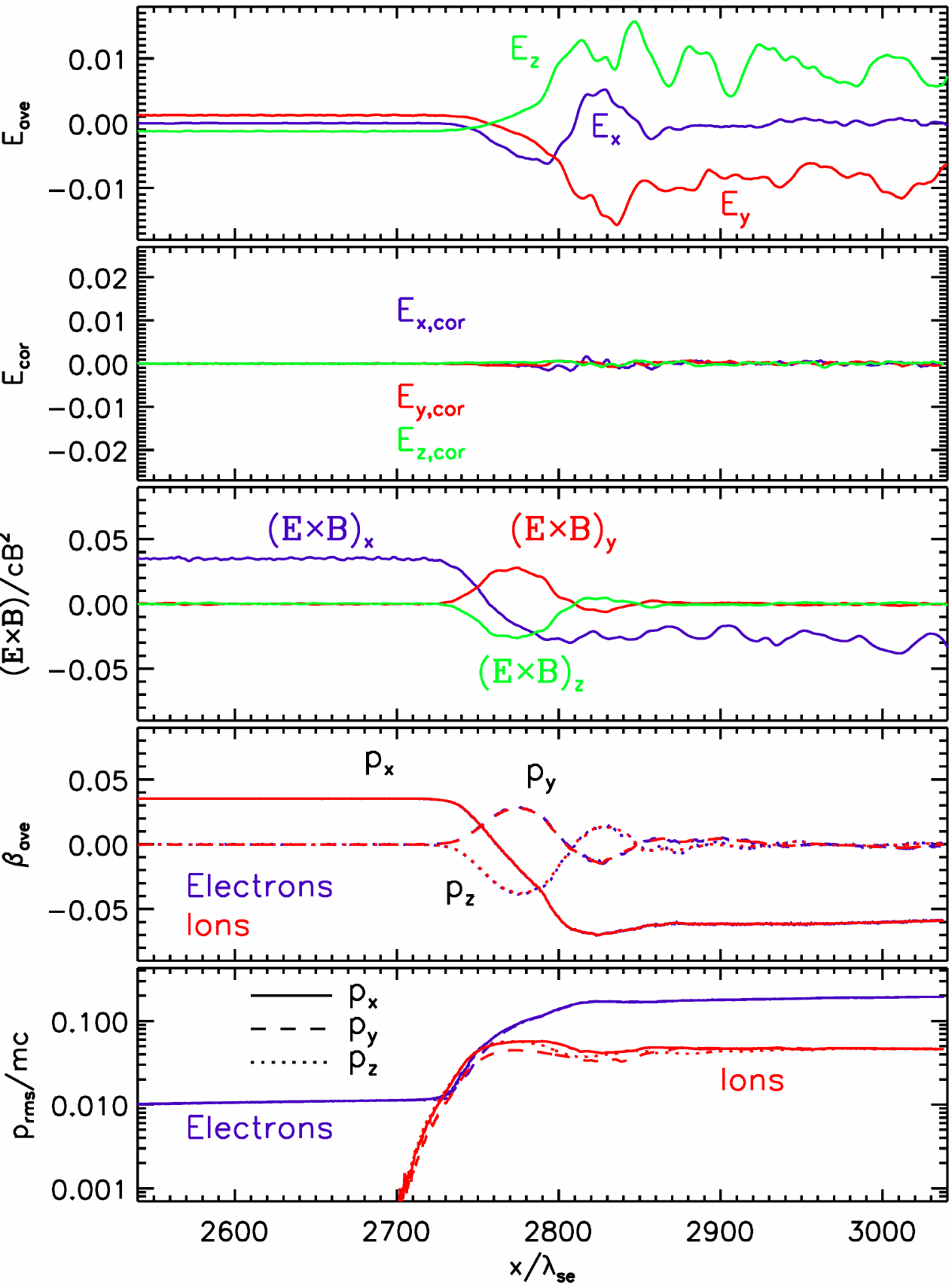}
\caption{Moments of the phase-space distribution of electrons and ions in comparison to the electric-field components, {both in the simulation frame and in the guiding-center frame,} and the components of $\mathbf{E}\times\mathbf{B}$ drift in the region around the reverse shock. All quantities are averaged over the $y$-coordinate. For ease of comparison with the drift speed we plot the average motion as velocity $\beta=p/mc(1+p^2)$.}\label{moments_rv}
\end{figure}
Figure~\ref{moments_rv} displays moments of the phase-space distributions of particles near the reverse shocks. To be noted is that particle drift along the shock surface is well described as $\mathbf{E}\times\mathbf{B}$ drift, as is the case of the forward shock, {and likewise the electric field is of very low amplitude in the guiding-center frame.} Gradient drift is a prerequisite of SDA but seems to play a negligible role in the dynamics of electrons and ions.

Fourier spectra of the electron density in the upstream region taken from the region $2660\,\lambda_\mathrm{se}$ -- $2725\,\lambda_\mathrm{se}$ indicate a mode at $k_\perp\,\lambda_\mathrm{se}\approx 0.8$ and nothing else. A corresponding signal is seen in $B_z$ at $k_\perp\,\lambda_\mathrm{se}\approx 0.6$. The same perpendicular mode was observed upstream of the forward shock (see Fig.~\ref{fourier}), albeit at slightly smaller scales. The parallel mode observed at the fast shock is absent ahead of the slower shock, at least in the electron density. In $E_x$ we see a parallel mode at $k_\parallel\,\lambda_\mathrm{se}\approx 6$ and an oblique mode similar to that visible in Figure~\ref{fourier} but at twice the wavenumber. Assuming that the parallel mode in $E_x$ is a Buneman wave, we can use Equation~\ref{eq0} to infer the drift velocity between electrons and ions to be $v_\mathrm{rel}\simeq 0.16\,c$, which corresponds to 30\% more than the shock speed, about the expected value. 

The Buneman mode is very weak far upstream, but in fact dominates the Fourier spectrum of $E_\parallel=\mathbf{E}\cdot \mathbf{e}_k$ in the foot region. We have to select small regions for further analysis on account of the large amplitude of shock rippling. For example, in the region $2720\,\lambda_\mathrm{se} \le x\le 2760\,\lambda_\mathrm{se}$ and $y\le 40\,\lambda_\mathrm{se}$ we see a strong signal in $E_\parallel=\mathbf{E}\cdot \mathbf{e}_k$ at $k_\parallel\,\lambda_\mathrm{se}\approx 6$ and $k_\perp \,\lambda_\mathrm{se}\le 3$. The phase-space distribution of particles in the same region indicates a drift of reflected ions relative to the electrons with $\mathbf{v}_\mathrm{rel}\simeq (-0.14\,c,0.07\,c)$, and so the relative speed is indeed $0.15\,c$ and the electrostatic mode can be identified as Buneman wave. 

Throughout the ramp we observe filaments in $\mathbf{j}$ with corresponding filamentary structure in the components of the magnetic field. Our analysis of the forward shock demonstrated that the current filaments are carried by incoming particles and are bent according to their deflection in the ramp region. Qualitatively, we see the same at the reverse shock. It is difficult to establish a quantitative match, though, because the strong shock rippling imposes significant variations in the local flow velocity of incoming particles. 
{Overall, the main structural difference between the forward and the reverse shock are caused by the existence of rippling, that are here fully resolved for the reverse shock only.}

\subsection{Particle Distributions}\label{distributions}
In a collisionless environment (and in our PIC simulation), isotropization and relaxation to near-Maxwellian distribution functions is achieved through interactions with plasma turbulence \citep{2015arXiv150200626B}, which can be described as a collision term in the Boltzmann equation \citep{2009PhRvL.102x5005B,2010PhPl...17e5704B}. At collisionless shocks this turbulent isotropization provides validity to a hydrodynamical description on large scales. High-energy tails in particle spectra may develop at a shock, though, that permit the injection of particles into Fermi-type acceleration processes. Care must be exercised to distinguish true spectral tails from apparent spectral structure that arises from incomplete isotropization.

\begin{figure}[htb!]
\centering
\includegraphics{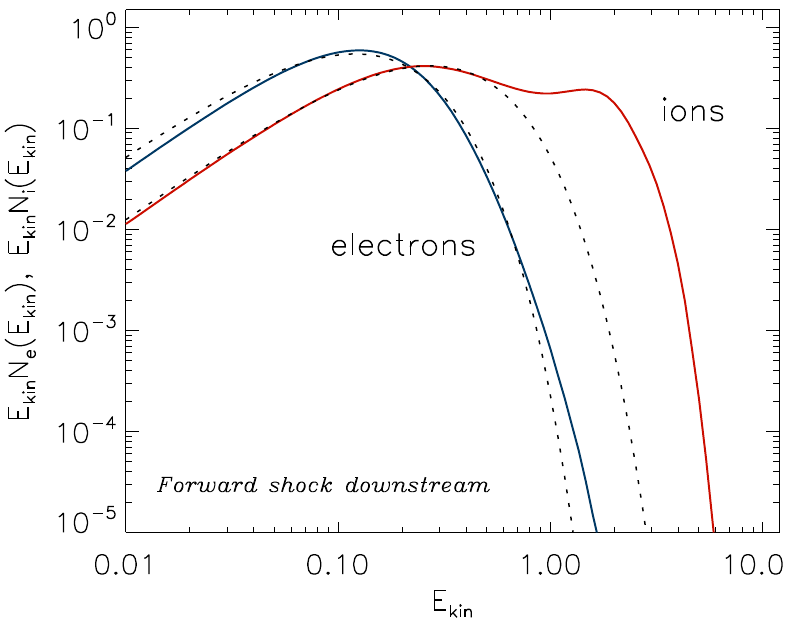}
\caption{Kinetic-energy spectra of electrons (blue line) and ions (red line) at time 
$t=20\,\Omega_\mathrm{i}^{-1}$ and the region $x\approx 5000-5700\,\lambda_\mathrm{se}$ far downstream of the forward shock at 
$x\approx 6800\,\lambda_\mathrm{se}$. The spectra are calculated in the downstream rest frame. They are normalized and expressed in simulation units, in which $m_ec^2=0.25$. The dotted lines indicate fits of a relativistic Maxwellian.}
\label{dist_r_shock}
\end{figure}

\begin{figure}[htb!]
\centering
\includegraphics{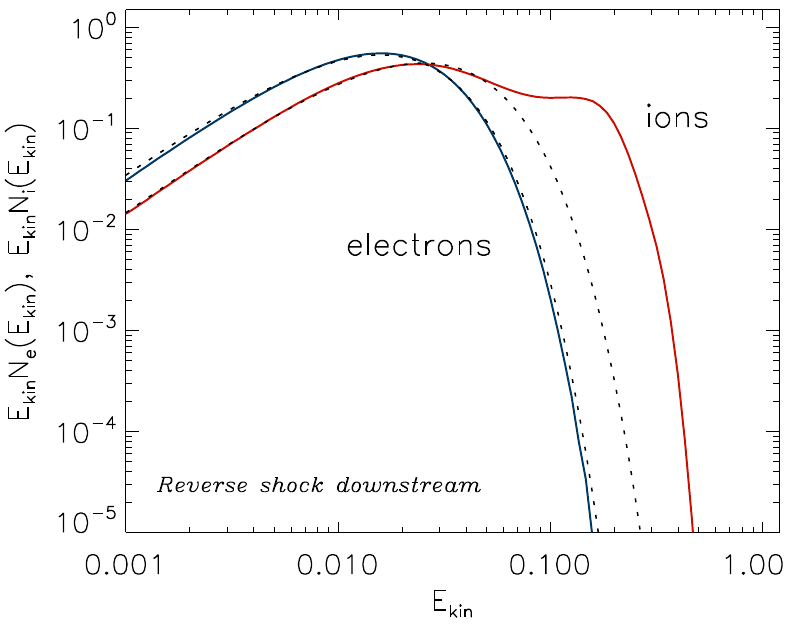}
\caption{Kinetic-energy spectra of electrons (blue line) and ions (red line)at time 
$t=20\,\Omega_\mathrm{i}^{-1}$ and in the region $x\approx 2910-2980\,\lambda_\mathrm{se}$ far downstream of the reverse shock at 
$x\approx 2800\,\lambda_\mathrm{se}$. The spectra are calculated in the downstream rest frame. They are normalized and expressed in simulation units, in which $m_ec^2=0.25$. The dotted lines indicate relativistic Maxwellian fits.}
\label{dist_l_shock}
\end{figure}

Figures \ref{dist_r_shock} and \ref{dist_l_shock} show particle distributions in kinetic energy, 
$E_\mathrm{kin}=(\gamma -1 )m_\mathrm{l}c^2 \;(m_\mathrm{l}=m_\mathrm{e}, m_\mathrm{i})$ downstream of the forward and the reverse shock, respectively, at $t=20\,\Omega_\mathrm{i}^{-1}$. The spectra are calculated in the downstream rest frame, defined by the CD speed $\beta_\mathrm{CD}=-0.06$ in the simulation frame.
We selected slices in the downstream region of the shocks that are about $5$ ion gyro radii in width considering the shock-compressed field, or about two ion gyro radii for the far-upstream magnetic field, $2\,v_x\,\Omega_\mathrm{i}^{-1}$. For the forward shock this implies a separation from the shock by $1100\,\lambda_\mathrm{se}$, whereas for the reverse shock the distance is only $110\,\lambda_\mathrm{se}$ on account of the much smaller shock speed. At these locations the plasma is homogeneous and free of particles that penetrated the contact discontinuity. 
In the CD rest frame, both ion and electron distributions are isotropic and display no large-scale bulk motion. 

For both the forward and the reverse shock, the distributions of electrons appear to be quasi-thermalized and well reproduced by relativistic Maxwellians. A Maxwellian fit to the bulk of the ion distributions suggests a temperature ratio $T_\mathrm{i}/T_\mathrm{e}\approx 2$. In addition, the ion spectra exhibit a significant supra-thermal tail. With the exception of electrons downstream of the reverse shock, whose temperature slightly increases with distance from the shock, particle spectra do not show significant evolution. Their distributions at fixed distance from the shock are also approximately steady in the time for $t\approx 5-20\,\Omega_\mathrm{i}^{-1}$, which demonstrates that the system has attained a quasi-equilibrium within the runtime of our simulations.

It is well known that the formation of supra-thermal tails in the ion distribution at perpendicular shocks can result from shock-surfing acceleration (SSA), in which shock-reflected particles performing half a gyro cycle upstream increase their energies through motion along the convective electric field. The maximum ion energy is roughly constant in our simulation, which suggests that a typical particle needs only a few gyro cycles to traverse the shock potential and be transmitted downstream. The pre-accelerated ions may be subject to further energization by DSA, a process that operates on larger scales than are covered in our simulations. However, recent hybrid simulations suggest a low DSA efficiency for perpendicular shocks \citep{capspit_2014a}.

The downstream electron distributions show either no (reverse shocks) or only marginal (forward shock) non-thermal populations. This is in contrast to the results of \citet{amahosh_2012} for their Runs A and C with parameters similar to ours, in particular a quite low plasma beta 
$\beta_\mathrm{p}$, where efficient acceleration by an electron SSA process was observed. This process 
requires nonlinear growth of the Buneman modes in the shock foot.  
Electrons in this region can be trapped in coherent electrostatic-potential structures and be directly accelerated by the convective electric field. They eventually escape from the potential wells and start to drift downstream. However, if a particle gains enough energy at the first encounter with the Buneman waves and its gyro radius increases, it can enter the electrostatic wave region again from the downstream side, and experience another SSA cycle \citep[see also][]{amahosh_2009}. 
As shown in Section~\ref{late_stage}, Buneman-type turbulence is generated in the foot of both the forward and the reverse shock, and in the former it grows to nonlinear amplitudes. 

We have seen during the early evolution of the simulation (see section~\ref{early_stage}), that the estimates of electron-acceleration efficiency presented in \citet{amahosh_2012} are too optimistic. It appears that fast electrons violate the trapping condition. Efficient electron heating can occur, as demonstrated in Sections~\ref{early_stage} and \ref{late_stage}. However, relativistic particles easily escape the electrostatic potential well and their acceleration is not efficient. 

Inefficient production of non-thermal electrons was also reported by \citet{kato_2010} in their simulations of an extremely high-Mach-number shock, $M_\mathrm{A}\sim 130$. As in the case studied here, shock formation was mediated by a Weibel-like filamentation instability, that generated mostly magnetic turbulence. Although an unstable electrostatic mode was found in the foot of the shock and identified as the Buneman wave, the growth of the instability was very slow due to the high temperature of the reflected ions. It is unclear what role is played by filamentation that may prohibit efficient growth of the Buneman waves to strongly nonlinear amplitudes, as similar filaments are seen by us and by, e.g., \citet{2013PhRvL.111u5003M}. The turbulent structure of the shock transition may also play an important role in impeding a return of electrons to the foot region or suppressing shock drift acceleration of electrons in the shock ramp, that should accompany the electron SSA process \citep{amahosh_2012}. 

\citet{kato_2010} attributed the lack of efficient electron acceleration to the specific orientation of the upstream magnetic field in the plane of the simulation, that was different from the out-of-plane configuration assumed in \citet{amahosh_2009}. 
The simulations by \citet{amahosh_2012} and \citet{2013PhRvL.111u5003M} with out-of-plane magnetic field and 
moderate plasma beta, $\beta_\mathrm{p}=0.5$, show high efficiency of electron energization. Little electron acceleration was seen by us for small plasma beta, {by \citet{amahosh_2012} for high $\beta_\mathrm{p}=4.5$}, and by \citet{kato_2010} for {very} high plasma beta, $\beta_\mathrm{p}\sim 26$, suggesting that the plasma beta is not the decisive factor. 


{The ion-to-electron mass ratio used in our study should also \emph{not} play a role. Although the publications by \citet{amahosh_2012} and \citet{2013PhRvL.111u5003M} are based on a larger $m_i/m_e$ of 100--225 compared to 30--50 as in 
\citet{kato_2010} and this paper, earlier studies by \citet{amahosh_2009} and also Run A in \citet{amahosh_2012} with $m_i/m_e=25$ show effective electron pre-acceleration through the SSA process. Larger ion-to-electron mass ratios lead to higher amplitudes of the magnetic field at the overshoot and in conditions allowing for efficient electron SSA enable further acceleration through adiabatic processes at the shock for already relativistic pre-accelerated electron populations \citep{amahosh_2012}.

The obvious difference between simulations showing efficient electron acceleration and studies yielding little electron energization is the configuration of the regular magnetic field component with respect to the simulation plane.}
It appears that strong acceleration {through SSA} is seen when the large-scale magnetic field is assumed to be strictly out of plane, as in \citet{amahosh_2009,amahosh_2012,2013PhRvL.111u5003M}. {On the other hand, turbulent} reconnection is seen only in a simulation with in-plane magnetic field \citep{2015Sci...347..974M}. 
{The Buneman modes have wavevectors preferentially aligned with the reflected-ion beam and thus are better resolved in 2D studies when the ion gyro-motion is contained in the plane of a simulation, i.e., for out-of-plane magnetic field configurations \citep{riqspit_2011}. Our simulation with $45^\circ$ field orientation should well resolve the Buneman instability and enable turbulent reconnection processes, thus allowing 3D physics to be approximated in a 2D simulation. The apparent lack of efficient electron acceleration in our setup may suggest that the efficacy of these kinetic instabilities in three dimensions is lower than in idealized 2D configurations. Fully three-dimensional simulations appear to be required.}

\subsection{Shock Reformation}\label{reformation}
Studies of low Mach number super-critical perpendicular shocks have demonstrated that the shock front 
is nonstationary and recurrently disappears and re-develops on a timescale of the order of the downstream ion gyroperiod. The process is called a cyclic self-reformation of the shock and is caused by the dynamics of the shock-reflected ions \citep[see, e.g.,][]{treumann2009,umeda2010,umeda2014}. 

\begin{figure}[htb!]
\centering
\includegraphics{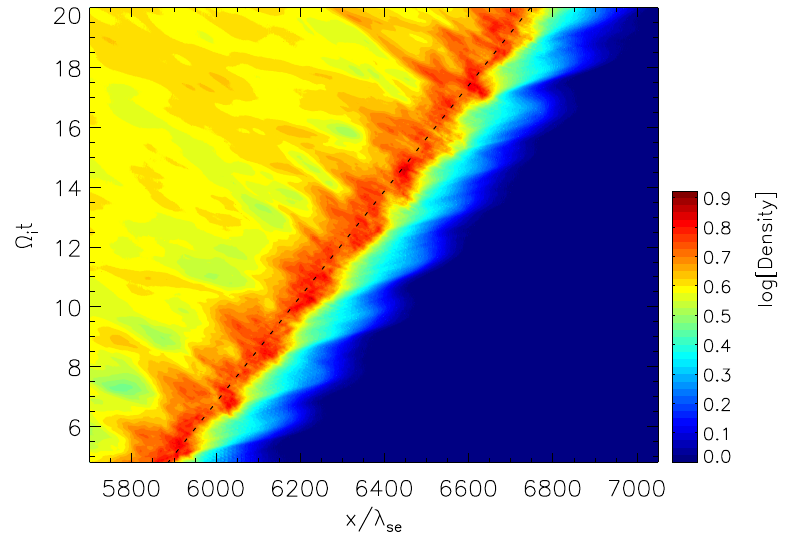}
\caption{Electron density in the vicinity of the forward shock as a function of position and time, averaged over the $y$-coordinate. The dashed line marks the mean location of the shock front propagating with $v_\mathrm{sh,R}$.}
\label{prof_aveR}
\end{figure}

Figure~\ref{prof_aveR} shows the time evolution of the average density profile in the vicinity of the forward shock. 
In this representation the shock moves from left to right with an average speed of $v_\mathrm{sh,R}$, as marked with the dashed line.
Although the shock never completely disappears a self-similar cyclic evolution of the average profile is evident. Self-repeating reformation phases occur with a period of approximately $1.5\,\Omega_\mathrm{i}^{-1}$ 
and are marked by 
shifts of the shock ramp position with the
enhancements in plasma density at the shock, and extensions of the filamentary region in the shock foot. 

\begin{figure*}[htb!]
\centering
\includegraphics[width=0.46\linewidth]{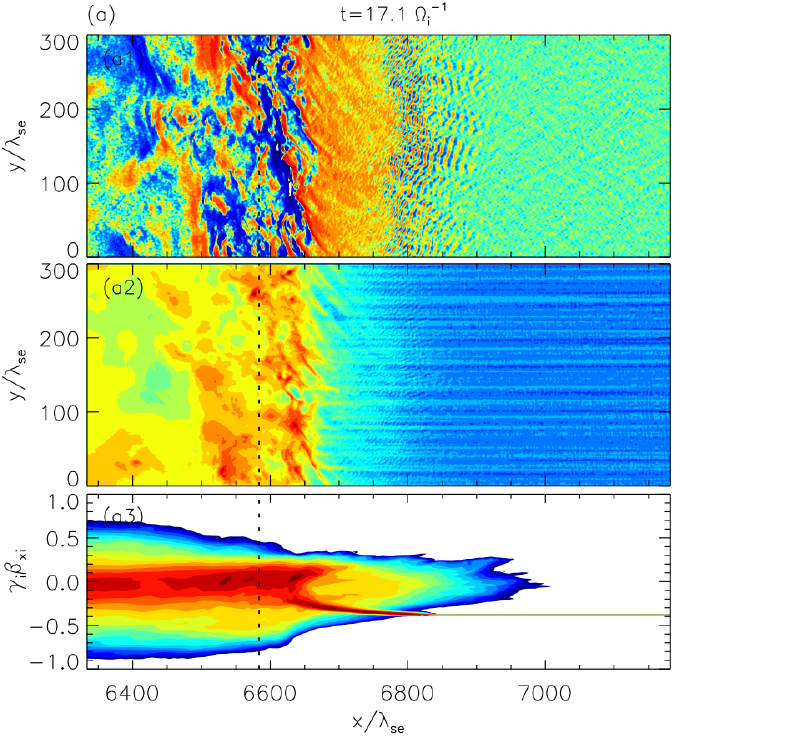}\hspace*{-0.7cm}
\includegraphics[width=0.46\linewidth]{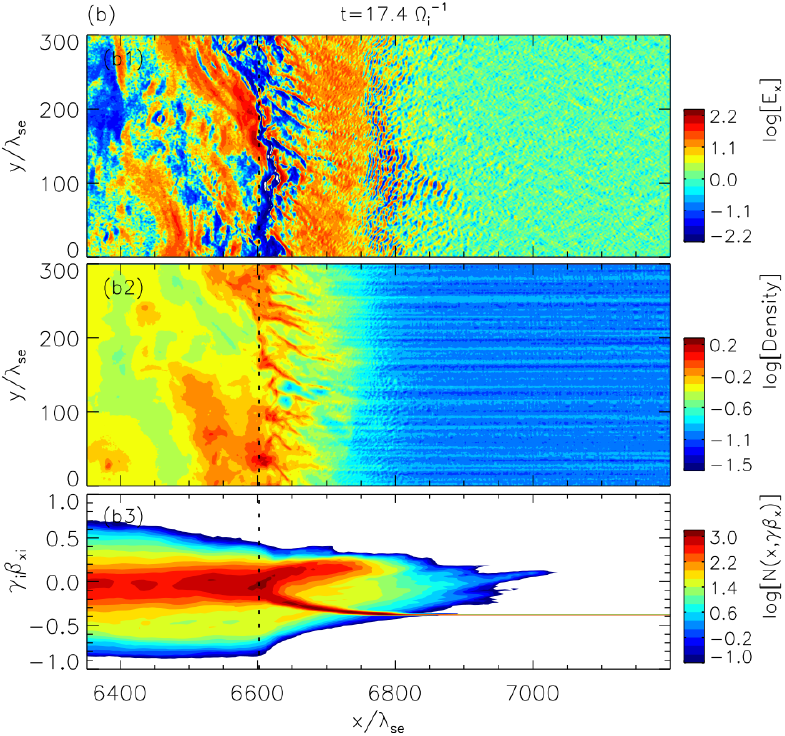}\\
\includegraphics[width=0.46\linewidth]{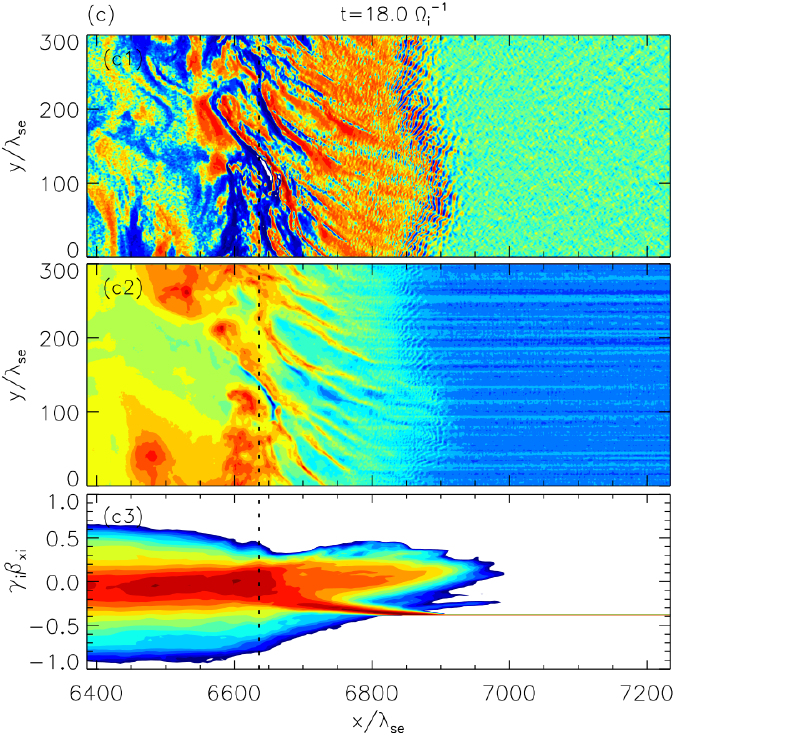}\hspace*{-0.7cm}
\includegraphics[width=0.46\linewidth]{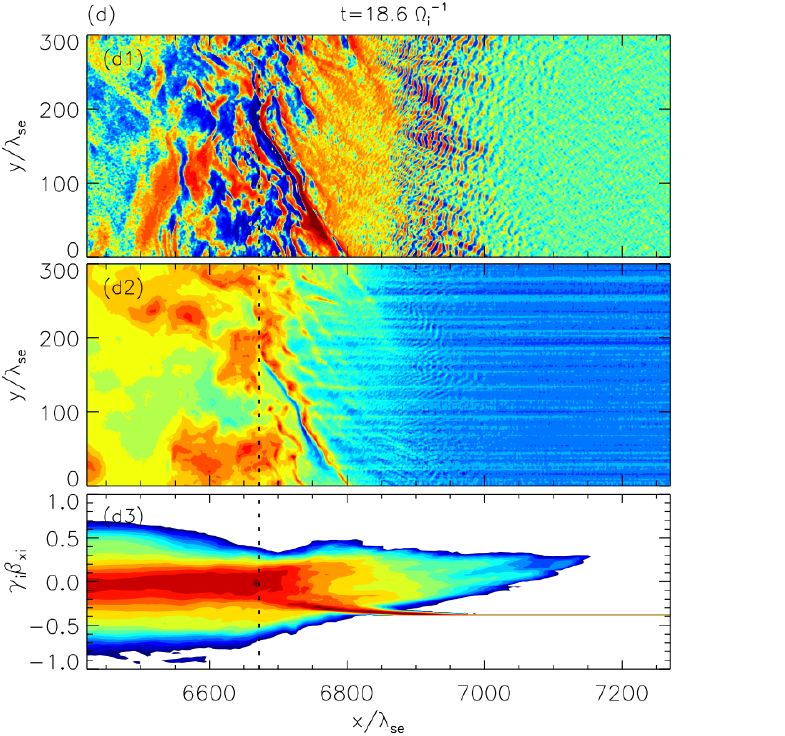}
\caption{Structure of the forward shock in a series of snapshots at times (a) $t=17.1\,\Omega_{i}^{-1}$, (b) $17.4\,\Omega_{i}^{-1}$,
(c) $18\,\Omega_{i}^{-1}$ and (d) $18.6\,\Omega_{i}^{-1}$, illustrating the main phases of a shock self-reformation cycle.
Displayed are from top to bottom in each panel the distributions of the electric-field component $E_x$, the density of electrons, and the longitudinal phase-space distribution of ions. The shock rest frame is used and the mean shock-ramp position is marked with a dashed vertical line.}
\label{reform}
\end{figure*}

The shock reformation arises because specular ion reflection from the shock ramp is not a continuous process. 
As frequently observed in low Mach number shocks, reflected ions bunch in the upstream edge of the shock foot due to either a non-steady reflection rate, or steepening of plasma waves excited in this region \citep[see, e.g.,][]{umeda2010}
In time the cross-shock potential builds at this location and a new shock front develops. In our simulation of high Mach number shocks we see similar effects. However, the physics of reflected ion bunching is now governed by the dynamics of current filaments resulting from the Weibel-like instability that mediates the shock transition.

The main features of a shock-reformation cycle in the forward shock are illustrated in Figure~\ref{reform} in a series of snapshots displaying distributions of the electron density, the $E_x$ electric field component, and the ion phase space. Results are displayed in the shock rest frame, and a mean shock ramp position that moves with a constant velocity of the forward shock $v_\mathrm{sh,R}$ is marked with a dashed vertical line (compare Fig.~\ref{prof_aveR}).  
At time $t=17.1\,\Omega_\mathrm{i}^{-1}$ (Fig.~\ref{reform}a) density compressions and associated quasi-regular electric field potential structures are formed
approximately $50\,\lambda_\mathrm{se}\simeq 2\,\lambda_\mathrm{si,R}$ from the shock ramp. There is no strong reflected ion beam in the foot region and weak current filaments are confined close to the shock front. This stage marks the beginning of a new shock-reformation cycle. 
By $t=17.4\,\Omega_\mathrm{i}^{-1}$ a strong beam of reflected ions develops with $v_x\approx -v_{\mathrm{R},x}$ and flow-aligned plasma filaments appear in the shock foot (Fig.~\ref{reform}b). 
As the beam traverses the foot region the size of the filamentary precursor increases and reaches its maximum extension at 
$t\approx 18\,\Omega_\mathrm{i}^{-1}$ (Fig.~\ref{reform}c). 
At the same time a location of maximum ion reflection at a position of the mean cross-shock potential moves with the average plasma flow towards the shock ramp. On the shock-ramp crossing the shock potential structures become disrupted and converge with the downstream plasma flow.
Current filaments start to merge ahead of the shock ramp which greatly enhances bipolar electric fields associated with them. 
{Once the gyration of ions} 
in the shock foot forces the filaments to align closer with the shock surface, the electric field structures begin to play a role of new shock potential fields. 
Efficient ion reflection thus sets off in the region extending up to $\sim 3\,\lambda_\mathrm{si,R}$ from the shock ramp and continues until reflected ion bunch becomes dispersed in the incoming plasma flow. By that time ($t\approx 18.6\,\Omega_\mathrm{i}^{-1}$, Fig.~\ref{reform}d) further filament mergers produce a new large-scale quasi-coherent electric field potential structure at $\sim 2\,\lambda_\mathrm{si,R}$ from the shock ramp and a new shock-reformation cycle begins.

The simplified picture of shock self-reformation delineated above may be further modified by local effects resulting from a non-coherent evolution of current filaments across the shock. Phases of the reformation cycles thus change along the shock surface and partially cancel out in the average profiles of Figure~\ref{prof_aveR}, which makes the reformation irregular. Moreover, the shock reformation process can also be influenced by the shock-front ripples. The latter was shown 
to modify the period of the reformation cycles in later stages in low Mach number perpendicular shocks \citep[e.g.,][]{2007GeoRL..3414109H,2009JGRA..114.3217L,umeda2010}. However, the rippling of the shock surface is weak in our case of the high Mach number forward shock and we do not see any apparent change of the reformation period throughout the duration of our simulation.

Finally, we note that cyclic self-reformation is also observed in the reverse shock. The physical nature of the process is the same as in the case of the forward shock. It is strongly influenced by the shock rippling, and can be visible only in the time evolution of the \emph{local} electron density profiles \citep[see, e.g.,][]{umeda2010}. Also in this case, a quasi-regular period of approximately $1.4\,\Omega_\mathrm{i}^{-1}$ does not change during the system evolution followed in our simulation.

\section{Summary and Discussion}\label{summary}
We have performed 2D3V PIC simulations of non-relativistic plasma collisions with perpendicular large-scale magnetic field. With the current study, we continue our investigations of the high-speed nonrelativistic shocks that started with the parallel shock simulations published in \citet{niemiec_2012}. Here, the main question we addressed was how changing the magnetic field from the parallel to perpendicular configuration influences the resulting shock structure and the efficiency of particle pre-acceleration.
{By configuring the large-scale magnetic field at an angle of $45^\circ$ to the simulation plane we expect to approximate the three-dimensional physics in a 2D system.}

To address this question, we 
{use a flow-flow setup with}
asymmetric plasma flows, i.e., utilizing the collision of plasma slabs of different density, leading to two different shocks and a CD that is self-consistently modeled. In contrast to the setup used by \citet{murphydieck_2010,murphydieck_2010a}, we avoid the creation of an artificial dipole {antenna} at the CD by using a transition zone between full motional electric fields inside the plasmas and zero electric field in a small plasma-free area between the plasma flows at the beginning of the simulation. Thereby, {with this new method} we ensure that our simulation is as clean as possible.

The simulation parameters are chosen such that they are close to those of a young supernova remnant, {and} we use a reduced ion-to-electron mass ratio of $m_i/m_e=50$ to capture both electron and ion dynamics within the bounds of our simulation. The density ratio between the dense plasma and the tenuous plasma is 10, which is about the same ratio we would expect to find at a young supernova remnant at the beginning of its free expansion phase in a typical ISM environment. However, shock velocities we normally find in such an environment are about a factor of a few lower than in our simulations. The sonic and Alfv\'{e}nic Mach numbers of the forward shock are $M_{s,R}=755$ and $M_{A,R}=27.6$, 
respectively, and that of the reverse shock are 
$M_{s,L}=252$ and $M_{A,L}=28.5$, 
respectively. Here, as forward and reverse shock we denote the shocks propagating into the tenuous plasma and into the dense plasma, respectively.

Our results can be summarized as follows:
\begin{enumerate}
\item Our newly developed setup leads to the creation of a very clean perpendicular shock without artificial transients that may limit the veracity of the simulation. Eventually, a double-shock structure evolves within a few ion cyclotron times, $\Omega_\mathrm{i}^{-1}$. The shock transition is mediated  by Weibel-type filamentation instabilities that lead to the development of current filaments and magnetic turbulence. 
\item {Shocks that form in the system are non-stationary and cyclically self-reform. As in low-Mach-number shocks, shock reformation is driven by a non-steady ion reflection.} The period of reformation is similar at both shocks with $\sim 1.5\,\Omega_\mathrm{i}^{-1}$. Generally, the periodicity of ion reflection is locally modified by the non-coherent evolution of current filaments across the shock.
\item The surface of {the reverse} shock is clearly rippled {on spatial and temporal scales given by the ions reflected at the shock. The ripples result from a modulation of the fraction of the shock-reflected ions along the shock surface, as in the scenario described by \citet{burgess-scholer_2007} for low-Mach-number shocks. Spatially-modulated ion reflection at the shock should lead to enhanced localized electron heating and acceleration.
The same instability operates in the forward shock, but its development is influenced by our boundary condition in the direction of the shock surface. We do not observe inertia-scale fluctuations that would arise, e.g., from the Alfv\'{e}n ion cyclotron instability driven by temperature anisotropy at the shock ramp. 
}
{The existence of ripples} affects the visibility of shock reformation, which is washed out in $y$-averaged density profiles of the reverse shock. 
\item We do not find any evidence for gradient drift at the shock. It is not possible to reconstruct from the available electromagnetic field data a uniform drift direction, because the local gradients in the {turbulent} magnetic field are much larger than the global gradient across the shock. We also do not find any indication of counterstreaming electrons and ions along the shock surface. In fact, the bulk motion of electrons and ions is commensurate with $\mathbf{E}\times \mathbf{B}$ drift in direction and amplitude. {To be noted is that the existence of an electric-field component parallel to the drift direction of particles does not imply the existence of a significant electric field in the drift frame. In fact, in the guiding-center frame given by the local bulk motion and $\mathbf{E}\times \mathbf{B}$ drift the electric field has a small amplitude, which explains the absence of electron acceleration in the ramp. For the ions, the guiding-center approximation is poorly justified, and so they may have trajectories along a significant electric field and be accelerated. }
\item Downstream of both shocks the electrons are well described by relativistic Maxwellians, suggesting that turbulence has been very efficient in relaxing the electron distribution function. Ion spectra are composed of a quasi-thermal bulk 
with $T_\mathrm{i}/T_\mathrm{e}\approx 2$ and a supra-thermal tail. We do not observe spectral variations with either distance from the shock or simulation time for $t\approx 5-20\,\Omega_\mathrm{i}^{-1}$, suggesting a quasi-equilibrium in the system. The supra-thermal ions appear to result from SSA. The constancy of the maximum ion energy suggests that their spending more than a few gyro cycles in the shock region is exceptional.
Electron heating arises from Buneman modes in the shock foot. Their amplitude is not high enough to prevent escape of relativistic electrons, and so we observe heating of the bulk as opposed to the creation of a spectral tail. There is no evidence of turbulent reconnection that was recently claimed to cause efficient electron energization \citep{2015Sci...347..974M}. The inefficient electron acceleration observed by us for low plasma beta and by \citet{kato_2010} for high plasma beta suggests that the plasma beta is not the deciding factor for the generation of a signicant non-thermal electron population. 
{The ion-to-electron mass ratio in the simulation does also not play a role, since efficient acceleration has been observed in simulations with much lower $m_i/m_e=25$, provided that suitable conditions for the nonlinear growth of the Buneman modes exist \citep{amahosh_2009,amahosh_2012}.}
There may be additional factors in the microphysics of high-Mach-number shocks mediated by filamentation that limit the amplitude of Buneman waves or prevent return of electrons to the foot region, resulting in a small number of supra-thermal electrons. 
{To this end} the configuration of the large-scale magnetic field in 2D3V simulations may have an impact, as recent simulations showing significant tails in electron spectra have an orientation strictly out of plane \citep{amahosh_2012,2013PhRvL.111u5003M}. 
{Our setup with $45^\circ$ orientation to the simulation plane should help suppressing such effects and allow 3D physics to be observed in a 2D simulation. The lack of efficient electron acceleration in our simulation suggests a lower efficacy of kinetic instabilities in three dimensions with respect to specific 2D configurations. Fully three-dimensional studies of electron pre-acceleration at high Mach number shocks are clearly needed.}
\end{enumerate}

\acknowledgments
The authors thank Jens Ruppel for his contribution in preparing the new setup for the perpendicular shock. This work used the Extreme Science and Engineering Discovery Environment (XSEDE), which is supported by National Science Foundation grant number ACI-1053575. The authors acknowledge the Texas Advanced Computing Center (TACC) at The University of Texas at Austin for providing HPC and visualization resources that have contributed to the research results reported within this paper. We are also grateful for HPC resources provided by The North-German Supercomputing Alliance (HLRN). V.W., M.P. and I.R. acknowledge support through grants PO 1508/1-1 and PO 1508/1-2 of the Deutsche Forschungsgemeinschaft. The work of J.N. is supported by Narodowe Centrum Nauki through research projects DEC-2011/01/B/ST9/03183
and DEC-2013/10/E/ST9/00662.
K.N. is supported by NSF AST-0908040, NNX12AH06G, NNX13AP-21G, and NNX13AP14G.

\bibliography{references}

\appendix
\section{Details on the Perpendicular Shock Setup}\label{setup_details}
To establish the transition zone, we taper off the magnetic field at the edge of both plasma slabs {while keeping the plasma density constant}. The full {homogeneous} magnetic field {of amplitude $B_0$} inside the left and right plasma {spans a range} from the left box boundary up to $x_{0L}$ and from $x_{0R}$ up to the right box boundary, respectively, while the transition zone extends from $x_{0L}$ to $x_{0L}+w_{\mathrm{grad}}$ and from $x_{0R}-w_{\mathrm{grad}}$ to $x_{0R}$, respectively. 
For the left transition zone we set: 
\begin{equation}\label{mag_grad_left}
\mathbf{B}_{\mathrm{L}}(x)=\dfrac{B_0}{2}\left[\cos\left(\dfrac{x-x_{0L}}{w_{\mathrm{grad}}}\pi\right)+1\right]\, (\cos(\phi)\mathbf{\hat{y}}+\sin(\phi)\mathbf{\hat{z}}),
\end{equation}
where $\mathbf{\hat{y}}$ and $\mathbf{\hat{z}}$ are unit vectors in $y-$ and $z-$direction 
{and we assume that the magnetic-field vector $\mathbf{B_{0}}$ lies in the $y-z$ plane at an angle $\phi$ to the $y-$axis.} 
For the right transition zone we have correspondingly:
\begin{equation}\label{mag_grad_right}
\mathbf{B}_{\mathrm{R}}(x)=\dfrac{B_0}{2}\left[\cos\left(\dfrac{x-x_{0R}}{w_{\mathrm{grad}}}\pi\right)+1\right]\, (\cos(\phi)\mathbf{\hat{y}}+\sin(\phi)\mathbf{\hat{z}}).
\end{equation}
The factor $1/2$ {serves to normalize} the term in brackets to values between 0 and 1. 
Outside the plasma slabs the magnetic field is $B_y=B_z=0$.

Note that whereas the magnetic field tapers off, the streaming velocity of the plasma slabs remains constant. The motional electric field in the transition zones thus follows the same profile as the magnetic field: 
\begin{equation}\label{el_grad_left}
\mathbf{E}_{\mathrm{L}}(x)=\dfrac{v_{\mathrm{L},x}B_0}{2}\left[\cos\left(\dfrac{x-x_{0L}}{w_{\mathrm{grad}}}\pi\right)+1\right]\, (\sin(\phi)\mathbf{\hat{y}}-\cos(\phi)\mathbf{\hat{z}}),
\end{equation}
and
\begin{equation}\label{el_grad_right}
\mathbf{E}_{\mathrm{R}}(x)=\dfrac{v_{\mathrm{R},x}B_0}{2}\left[\cos\left(\dfrac{x-x_{0R}}{w_{\mathrm{grad}}}\pi\right)+1\right]\, (\sin(\phi)\mathbf{\hat{y}}-\cos(\phi)\mathbf{\hat{z}}).
\end{equation}
\begin{figure*}[htb!]
\includegraphics[width=0.49\linewidth]{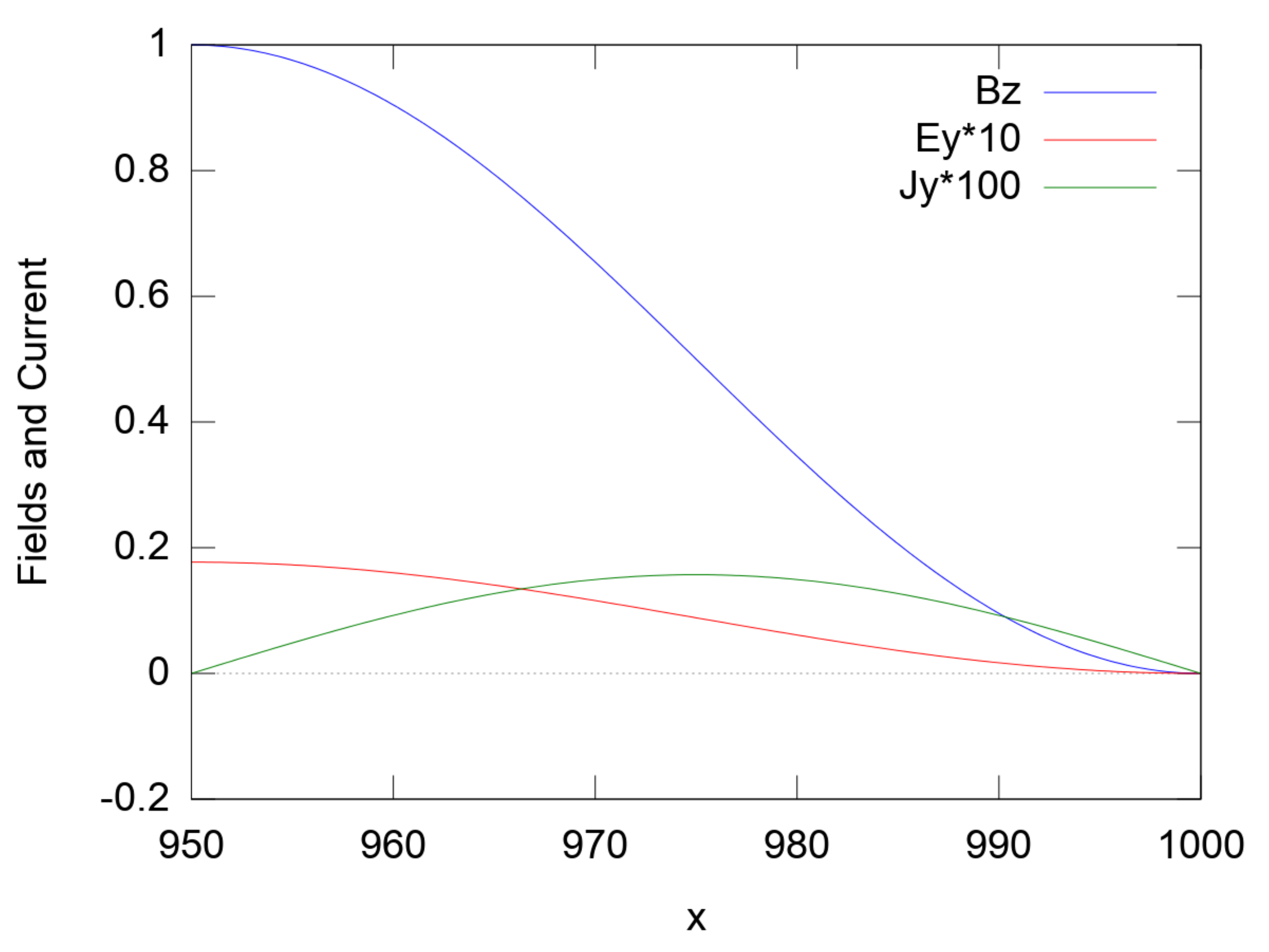}
\includegraphics[width=0.49\linewidth]{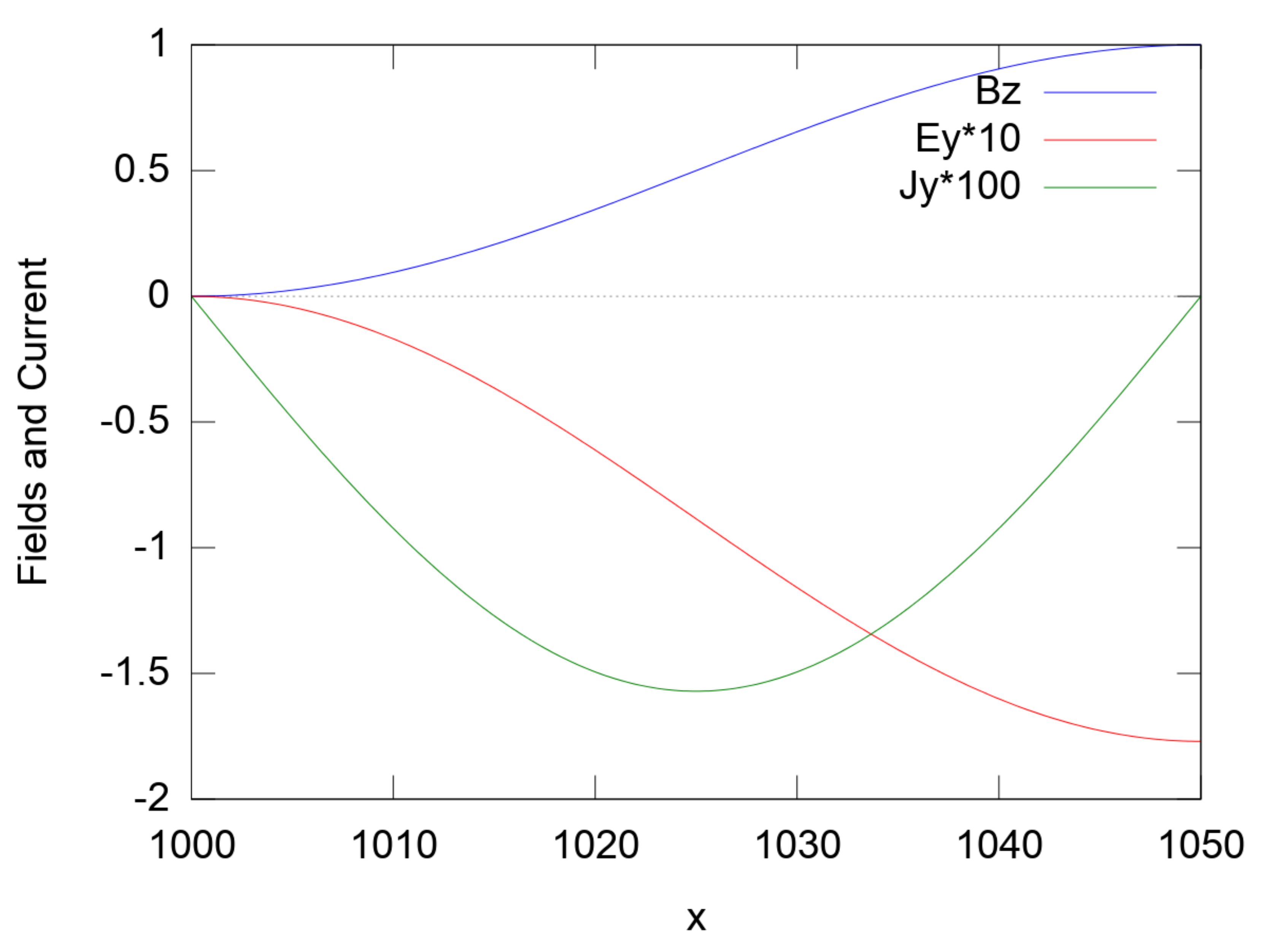}
\caption{These two figures illustrate the setup of the transition zone in the left (left panel) and in the right (right panel) plasma 
{for $\phi=90^{\rm{o}}$}. Shown are the perpendicular magnetic field $B_z$ (\textit{blue}), the motional electric field $E_y\times 10^1$ (\textit{red}) and the current sheet $J_y\times 10^2$ (\textit{green}).
{In this example, the width of the transition zone for both slabs is set to $w_{\mathrm{grad}}=50\Delta$ and there is no plasma-free area between  the beams.}}\label{gradient}
\end{figure*}
To achieve a stable gradient, one has to impose a current sheet in the transition zone that balances $\nabla\times\mathbf{B}$, so that in Heaviside-Lorentz units
\begin{equation}
\frac{\partial\mathbf{E}}{\partial t}=c\nabla\times\mathbf{B}-\mathbf{J}=0.
\end{equation}
As we have gradients only in $x$-direction, the current density $\mathbf{J}$ has non-zero components in $y-$ and $z-$direction, namely $J_y=-c\,\partial B_z/\partial x$ and $J_z=c\,\partial B_y/\partial x$. The current is set up through a drift of ions relative to the electrons, and as the particle density, $n$, is kept constant in the transition zone, $\mathbf{J}=n\,q\,\mathbf{v}_{\mathrm{rel}}$, the drift velocity $\mathbf{v}_{\mathrm{rel}}$ is calculated as:
\begin{equation}\label{vel_grad_left}
\mathbf{v}_{\mathrm{rel,L}}(x)=\dfrac{B_0\,c\pi}{2\,n_L\,q\,w_{\mathrm{grad}}}\sin\left(\dfrac{x-x_{0L}}{w_{\mathrm{grad}}}\pi\right)\, (\sin(\phi)\mathbf{\hat{y}}-\cos(\phi)\mathbf{\hat{z}}),
\end{equation}
and
\begin{equation}\label{vel_grad_right}
\mathbf{v}_{\mathrm{rel,R}}(x)=\dfrac{B_0\,c\pi}{2\,n_R\,q\,w_{\mathrm{grad}}}\sin\left(\dfrac{x-x_{0R}}{w_{\mathrm{grad}}}\pi\right)\, (\sin(\phi)\mathbf{\hat{y}}-\cos(\phi)\mathbf{\hat{z}}),
\end{equation}
where $n_L$ and $n_R$ are plasma particle densities in the left and the right plasma {slabs, respectively}.

\begin{figure*}[htb!]
\centering
\includegraphics[width=0.49\linewidth]{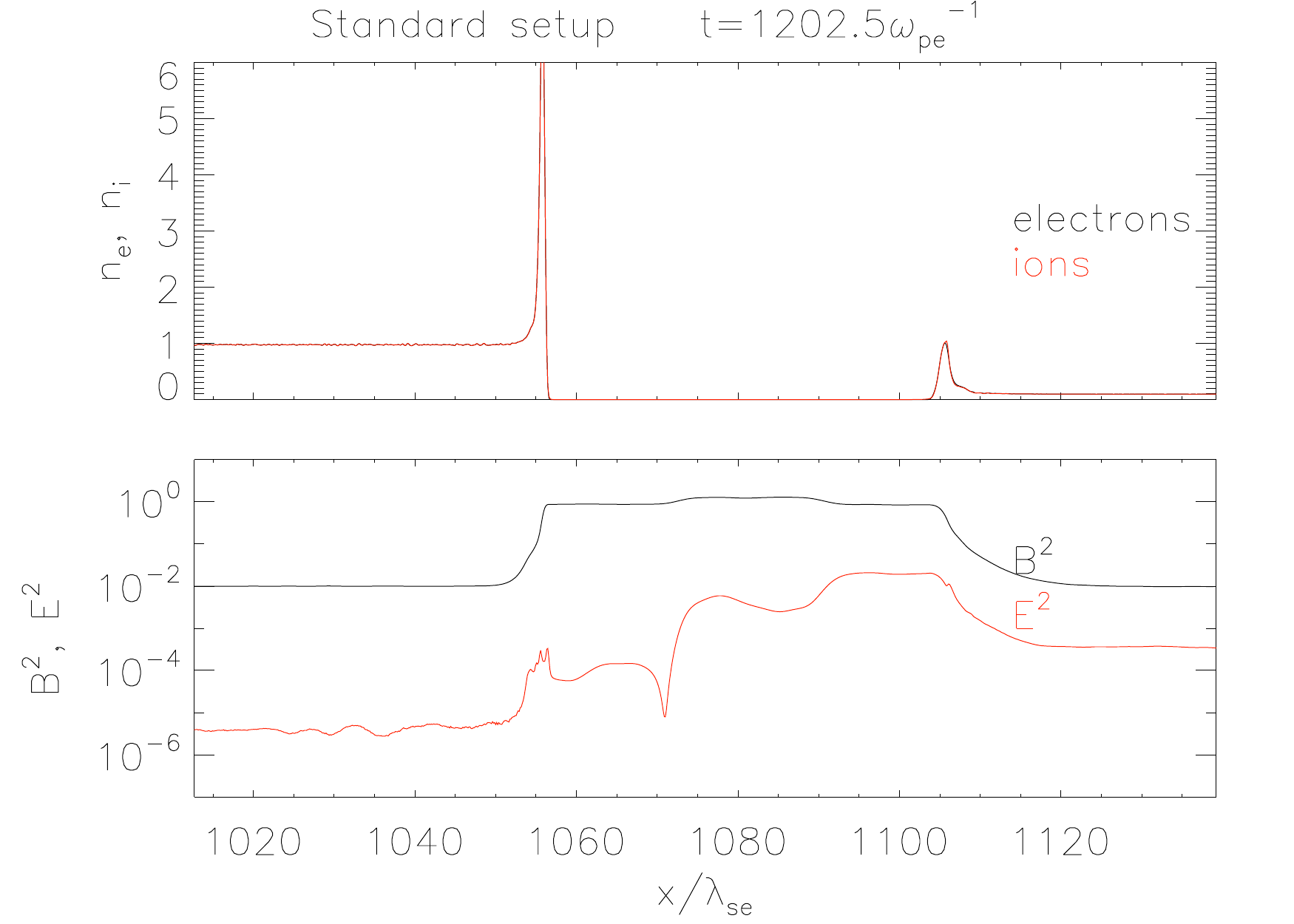}
\includegraphics[width=0.49\linewidth]{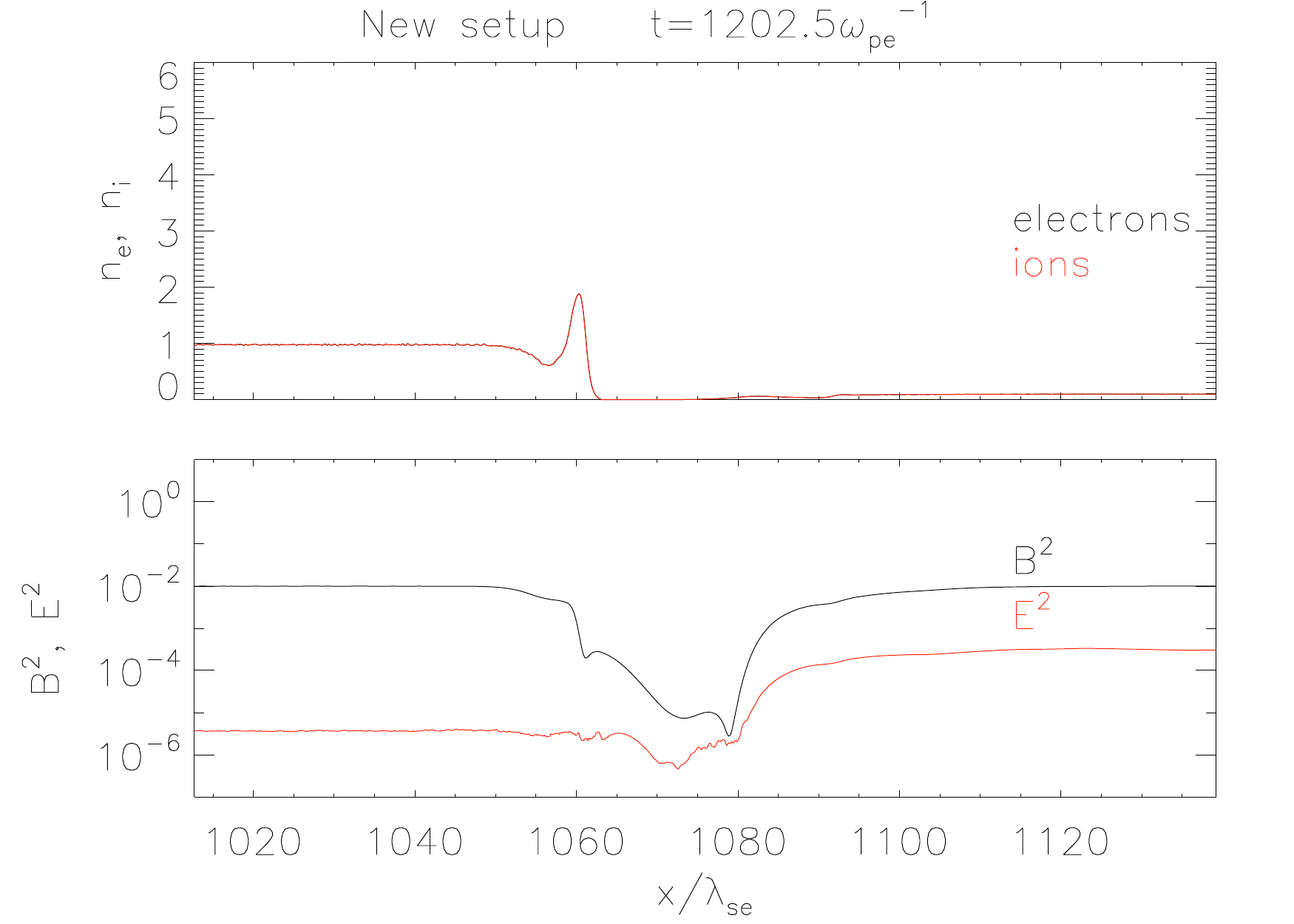}
\caption{Stability comparison of the standard setup (constant perpendicular magnetic field $B_z$ and jumping motional electric field $E_y$; left panel), and the new setup with the transition zone (right panel). Displayed is the state after $1202.5\,\omega_{pe}^{-1}$, shortly before the collision of the two plasmas. Particle densities are shown in the top panels, whereas {the lower panels display the field energy densities}. {In these demonstration simulations}, the collision {layer} is at $x\approx1063\,\lambda_{se}$ and 
{the two plasma slabs started with $v_{\mathrm{L},x}$ and $v_{\mathrm{R},x}$, respectively,}
at $x=1012.7\,\lambda_{se}$ and $x=1519.0\,\lambda_{se}$.}
\label{compare}
\end{figure*}

The current is carried by ions to guarantee sufficient stability of the current sheet. 
{For a suitable choice of $w_{\mathrm{grad}}$, the two plasmas fully collide after $2w_{\mathrm{grad}}/(v_{\mathrm{L},x}-v_{\mathrm{R},x})\approx 1\,\Omega_{e}^{-1}\approx 1/50\,\Omega_{i}^{-1}$, implying that the current carried by the ions is directionally stable on the timescale of the plasma collision}. We conducted tests to verify that no significant Buneman instabilities arise on account of the drift between electrons and ions.
Therefore, all instabilities observed in our simulation arise {solely} from the collision.
Figure \ref{gradient} illustrates the setup of the transition zone 
{for the out-of-plane perpendicular magnetic field orientation, i.e, $\phi=90^o$.} 

Figure \ref{compare} compares the stability in particle density and field energy density of the standard setup with a constant perpendicular magnetic field $B_z$ throughout the simulation box and the new setup with a transition zone as described above. One can clearly see, that the new setup is very stable over many time steps, and the area between the plasmas stays largely free of electromagnetic fields until the plasma slabs smoothly collide with each other.
In contrast, the standard method introduces strong transient fields and leads to strong density compressions 
{at fronts of the approaching plasma beams}
before {the onset of} the collision. One can observe a transient in the electric field, which is emitted at the location of the sign flip in $\mathbf{E}$ in the middle of the simulation box, and which eventually perturbs the magnetic field. 
{Note, that in a realistic setup the two plasma slabs are set much closer in the beginning than in this demonstration simulation; e.g., for the setup with $w_{\mathrm{grad}}=50\Delta$ shown in Figure~\ref{gradient} the plasmas} fully collide after about $32\,\omega_{pe}^{-1}$.

\end{document}